%% file: main.tex
\documentclass[10pt,conference]{IEEEtran}

\usepackage{amsmath,amsfonts}

\usepackage{hyperref}
\usepackage{subcaption}
\usepackage{adjustbox}
\newsavebox{\mybox}

\input{header}

\begin{document}

\title{Risk Estimation in Differential Fuzzing via \\ Extreme Value Theory} 

\author{
\IEEEauthorblockN{Rafael Baez}
\IEEEauthorblockA{
University of Texas at El Paso\\
rbaez2@miners.utep.edu
}
\\
\IEEEauthorblockN{Marcelo Frias}
\IEEEauthorblockA{
University of Texas at El Paso\\
mfrias4@utep.edu
}
\and
\IEEEauthorblockN{Alejandro Olivas}
\IEEEauthorblockA{
University of Texas at El Paso\\
aolivas23@miners.utep.edu
}
\\
\IEEEauthorblockN{Yannic Noller}
\IEEEauthorblockA{
Ruhr University Bochum\\
yannic.noller@acm.org
}
\and
\IEEEauthorblockN{Nathan K. Diamond}
\IEEEauthorblockA{
University of Texas at El Paso\\
nkdiamond@miners.utep.edu
}
\\
\IEEEauthorblockN{Saeid Tizpaz-Niari}
\IEEEauthorblockA{
University of Illinois Chicago\\
saeid@uic.edu
}

}

\IEEEtitleabstractindextext{
\begin{abstract}
Differential testing is a highly effective technique for automatically detecting software bugs and vulnerabilities when the specifications involve
an analysis over multiple executions simultaneously. Differential fuzzing, in particular, operates as a guided randomized search, aiming to find (similar) inputs that lead to a maximum difference in software outputs or their behaviors. However, fuzzing, as a dynamic analysis, lacks any guarantees on the absence of bugs: from a differential fuzzing campaign that has observed no bugs (or a minimal difference), what is the risk of observing a bug (or a larger difference) if we run the fuzzer for one or more steps?

This paper investigates the application of Extreme Value Theory (EVT) to address the risk of missing or underestimating bugs in differential fuzzing.
The key observation is that differential fuzzing as a random process resembles the maximum distribution of observed differences. 
Hence, EVT, a branch of statistics dealing with extreme values, is an ideal framework to analyze the tail of the differential fuzzing campaign to contain the risk.
We perform experiments on a set of real-world Java libraries and use differential fuzzing to find information leaks via side channels in these libraries.
We first explore the feasibility of EVT for this task and the optimal hyperparameters for EVT distributions. We then compare EVT-based extrapolation against baseline statistical methods like Markov's as well as Chebyshev's inequalities, and the Bayes factor.
EVT-based extrapolations outperform the baseline techniques in 14.3\% of cases and tie with the baseline in 64.2\% of cases. Finally, we evaluate the accuracy and performance gains of EVT-enabled differential fuzzing in real-world Java libraries, where we reported an average saving of tens of millions of bytecode executions by an early stop.
\end{abstract} 
}
\maketitle

\IEEEdisplaynontitleabstractindextext

\input{Sections/introduction}

\input{Sections/background}

\input{Sections/overview}

\input{Sections/problem-statement}

\input{Sections/approach}

\input{Sections/experiments}

\input{Sections/discussion}

\input{Sections/conclusion}


\bibliography{references}
\bibliographystyle{IEEEtran}

\end{document}

%% file: header.tex
\usepackage{amsthm}
\usepackage{mdframed}
\usepackage{centernot}

  \usepackage[defaultcolor=black]{changes}
\makeatletter
\setdeletedmarkup{\@gobble{#1}}
\colorlet{Changes@Color}{red}
\makeatother

\newcommand{\Aa}{\mathcal{A}}

\newcommand{\Hh}{\mathcal{H}}

\newcommand{\Pp}{\mathcal{P}}

\newcommand{\pP}{\mathbb{P}}

\newcommand{\Dd}{\mathcal{D}}

\newcommand\vd[2]{d_{i, p}}
\newcommand{\set}[1]{\left\{ #1 \right\}}

\newcommand{\dfuzz}{\textsc{DifFuzz}\xspace}
\newcommand{\diffuzz}{\textsc{DifFuzz}\xspace}
\newcommand{\blazer}{\textsc{Blazer}\xspace}
\newcommand{\themis}{\textsc{Themis}\xspace}
\usepackage{tikz}

\newtheorem{definition}{Definition}[section]

\definecolor{gold}{rgb}{0.99,0.78,0.07}
\usetikzlibrary{arrows,shapes,snakes,automata,backgrounds,positioning,decorations.pathmorphing,
	decorations.markings,calc}

\tikzstyle{dtreenode}=[draw=blue!10!gray,rounded rectangle, minimum size=5mm,fill=blue!10!white]
\tikzstyle{dtreeleaf}=[draw=black!60,minimum width=1cm,minimum height=0.4cm,rectangle,fill=blue!50!white]
\tikzset{every loop/.style={looseness=7}}
\tikzset{
	gluon/.style={decorate,draw=black,
		decoration={coil,amplitude=1pt, segment length=5pt}}
}
\tikzset{
	gluon1/.style={decorate,draw=black,
		decoration={coil,amplitude=3pt, segment length=3pt}}
}
\tikzset{
	gluonew/.style={decorate,draw=black,
		decoration={coil,amplitude=1pt, segment length=2pt}}
}

\tikzset{bicolor/.style args={#1 and #2 and #3}{
		path picture={
			\tikzset{rounded corners=0}
			\fill [#1] (path picture bounding box.south west)
			rectangle
			($(path picture  bounding box.north west)!#3!(path picture bounding
			box.north east)$);
			\fill [#2]
			($(path picture bounding box.south west)!#3!(path picture bounding
			box.south east)$)
			rectangle (path picture bounding box.north east);
}}}

\tikzset{tricolor/.style args={#1 and #2 and #3 and #4 and #5}{
		path picture={
			\tikzset{rounded corners=0}
			\fill [#1] (path picture bounding box.south west)
			rectangle
			($(path picture  bounding box.north west)!#4!(path picture bounding
			box.north east)$);
			\fill [#2]
			($(path picture bounding box.south west)!#4!(path picture bounding
			box.south east)$)
			rectangle
			($(path picture  bounding box.north west)!#5!(path picture bounding
			box.north east)$);
			\fill [#3]
			($(path picture bounding box.south west)!#5!(path picture bounding
			box.south east)$)
			rectangle (path picture bounding box.north east);
}}}

\usepackage[many]{tcolorbox}
\tcbuselibrary{listings,skins}

\lstdefinestyle{mystyle}{
  xleftmargin=0pt,
   basicstyle={\footnotesize\ttfamily},
   aboveskip=3mm,
   belowskip=3mm,
   keywordstyle=\bfseries,
   showstringspaces=false,
  escapechar=?,
  language=Java
}
\definecolor{code_indent}{HTML}{CCCCCC}

\newcommand{\indentrule}{\color{code_indent}\vrule\hspace{2pt}}

\newtcblisting{mylisting}[2][]{
  arc=0pt, outer arc=0pt,
  listing only,
  listing style=mystyle,
  title={\large #2},
  #1
}

 \definecolor{dkgreen}{rgb}{0,0.6,0}
 \definecolor{gray}{rgb}{0.5,0.5,0.5}
 \definecolor{mauve}{rgb}{0.58,0,0.82}

\definecolor{cadmiumgreen}{rgb}{0.0, 0.42, 0.24}
\definecolor{verde}{rgb}{0.25,0.5,0.35}
\definecolor{jpurple}{rgb}{0.5,0,0.35}
\definecolor{darkgreen}{rgb}{0.0, 0.2, 0.13}

\usepackage[ruled,vlined,linesnumbered,lined]{algorithm2e}
\usepackage{algpseudocode}

\usepackage{mathtools,xparse}

\usepackage{tabu}
\usepackage{multirow}

\usepackage{pifont}
\usepackage{enumerate}
\usepackage[shortlabels]{enumitem}

\usepackage{pgfplotstable}
\usepackage{pgfplots}

\usepackage[noframe]{showframe}
\usepackage{framed}

 {\endMakeFramed}
 \definecolor{shadecolor}{gray}{0.85}

\definecolor{bgblue}{RGB}{245,243,253}
\definecolor{ttblue}{RGB}{91,194,224}

\mdfdefinestyle{mystyle}{%
  rightline=true,
  innerleftmargin=10,
  innerrightmargin=10,
  outerlinewidth=3pt,
  topline=false,
  rightline=false,
  bottomline=false,
  skipabove=\topsep,
  skipbelow=false
}

\newtcolorbox{myboxi}[1][]{
  breakable,
  title=#1,
  colback=white,
  colbacktitle=white,
  coltitle=black,
  fonttitle=\bfseries,
  bottomrule=0pt,
  toprule=0pt,
  leftrule=3pt,
  rightrule=3pt,
  titlerule=0pt,
  arc=0pt,
  outer arc=0pt,
  colframe=black!50,
}

\newtcolorbox{myboxii}[1][style=mystyle]{
  breakable,
  freelance,
  colback=white,
  colbacktitle=white,
  coltitle=black,
  fonttitle=\bfseries,
  bottomrule=0pt,
  boxrule=0pt,
  colframe=white,
  after skip=0pt,
  overlay unbroken and first={
    \draw[white!75!black,line width=3pt]
    ([yshift=-9pt]frame.north west) --
    ([yshift=9pt]frame.south west);
  },
}

%% file: Sections/introduction.tex
\section{Introduction}
\label{sec:intro}
Modern software systems are notoriously prone to failures, which in 2020 alone caused an estimated \$1.56 trillion in economic losses in the U.S.~\cite{krasner2021cost}. To mitigate such risks, developers rely on debugging, testing, and verification as critical processes for identifying and preventing bugs. A major research and engineering challenge is to design automated techniques that make these processes scalable and effective. Among existing approaches, fuzzing has emerged as one of the most powerful techniques for automatically detecting bugs and vulnerabilities~\cite{bohme2020fuzzing}.

A widely used variant of fuzzing is graybox fuzzing~\cite{10.1145/3243734.3243804}. It employs evolutionary algorithms to search the input space, guided by program internals such as whether a test input explores a new path in the control-flow graph.
Despite its effectiveness, fuzzing can only demonstrate the absence of bugs for the specific inputs it generates, leaving uncertainty about untested inputs.
To address this uncertainty, there have been significant efforts within the software engineering community to provide statistical guarantees on the fuzzing campaign~\cite{bohme2018stads,bohme2020fuzzing,bohme2021estimating,reachability,saha2022preach,saha2023rare}. In particular, they aim to answer questions like: ``What is the likelihood of identifying a new vulnerability when running the fuzzing campaign for many more hours?'' 

Most existing work provides statistical guarantees of fuzzing, focusing on finding single inputs that witness a crash. However, there are significant classes of software requirements that cannot be tested via single input/output generation. Those include software correctness requirements that face the oracle problem~\cite{metamorphic,chen-original,tian2018deeptest,tizpaz2023metamorphic,tizpaz2020detecting,tizpaz2018differential} (i.e., the ground truth is not available), hyperproperties~\cite{8812124,noller2021qfuzz,tizpaz2019quantitative,tizpaz2019efficient,ruan2024timing,monjezi2025fairness} (i.e., two or more traces together deem correctness or witness a bug), and differential testing~\cite{mckeeman1998differential,petsios2017nezha,pei2017deepxplore} (i.e., finding bugs by comparing execution results of similar implementations).


%

Due to significant applications and challenges in the domain of differential analysis, for example, with differential fuzzing, we study the feasibility of statistical guarantees for this class of software testing methods. Differential testing represents a random process that draws random variables, each taking a value of outputs or behavior differences between two or more similar inputs when executed on the target software. Similar to the statistical guarantees in conventional fuzzing, we should be able to derive statistical claims about the maximum of such random variables in a given period of time even if the underlying distribution is unknown. Therefore, we pose a central question at the heart of differential fuzzing:

\begin{tcolorbox}[boxrule=1pt,left=1pt,right=1pt,top=1pt,bottom=1pt]
From a differential fuzzing campaign that has witnessed a difference $\delta \geq 0$ after $t$ iterations, what is the risk of observing larger differences if we run the fuzzer for one or more steps longer?
\end{tcolorbox}

We posit that the worst-case divergence of differential fuzzing represents the maximum of random variables in the fuzzing campaign. 
Therefore, the statistics of extremes~\cite{gumbel1958statistics} that model the tail distribution of random variables is a natural choice. Crucially, extreme value theory (EVT)~\cite{coles2001introduction} can reason about the likelihood as well as the amounts of extreme values (return levels) in a given period of time (return period). 
Hence, we pose the following research question: 

\begin{tcolorbox}[boxrule=1pt,left=1pt,right=1pt,top=1pt,bottom=1pt]
Given an observed (max) cost difference $\delta \geq 0$ between two (similar) inputs at iteration $m$ of differential fuzzing, can EVT estimate the maximum differences in the next $n$ iterations, up to an upper bound $n \leq N$?  
\end{tcolorbox}

If so, we could minimize the runtime of differential fuzzing campaigns by an early termination while still maintaining a high-quality in finding significant differences.
In this paper, we focus on the class of differential fuzzing for side-channel analysis. Specifically, we consider \textsc{DifFuzz}~\cite{8812124}, a state-of-the-art technique that searches for two secret inputs under the same public input, leading to a maximum cost difference in their executions on the target software.
Identifying such a high-cost difference between secrets indicates a vulnerability that can allow attackers to exploit timing side channels to infer information about the secrets. Note that small cost differences may not be manifested in the execution times; hence we require to find strong differences to deem a vulnerability. Therefore, \textsc{DifFuzz} is an ideal case study for our framework. 


However, there are multiple challenges in adapting EVT to provide statistical guarantees in differential fuzzing.
One challenge is the notion of statistical testing on the tail samples during the fuzzing to stop differential fuzzing.
While such notions are well studied for the statistics of regular distributions (e.g., normal distributions), an appropriate notion of statistical testing of tail distributions is challenging due to the scarcity. 
While the generalized extreme value (GEV)~\cite{leadbetter2012extremes} distributions (as a result of EVT theory) can reason about the tail, they require setting different parameters that can significantly influence the outcome. 
One such parameter is a threshold parameter that deems any sample that exceeds this threshold a tail sample.
The proper choice of threshold is critical. An underestimation of the (ground truth) threshold leads to mixture distributions that violate the asymptotic basis of GEV distributions. 
Similarly, an overestimation can only include fewer tail samples, which can lead to low confidence in the model due to high variance. 
There are also different types of GEV distributions, such as Exponential, Poisson Process, and Pareto Distributions, that have different statistical properties. 
So, the first question is:

\vspace{0.25 em}
\noindent \textbf{RQ1.} \emph{How to infer an ideal statistical testing of tail distribution and hyperparameters of GEV distributions as well as its type to accurately estimate the worst-case differential fuzzing?}

Once we infer suitable configurations of tail extrapolations, the next challenge is to establish their efficacy when compared to the baseline statistical techniques. One class of such techniques is based on concentration inequalities, such as Markov's and Chebyshev's inequalities. However, these techniques focus on the tails of regular distributions, rather than modeling the tail distribution directly. Another class of algorithms is various derivations of the Bayes factor that rely on statistical hypothesis testing based on the sequence of extreme events. 

\vspace{0.25 em}
\noindent \textbf{RQ2.} \emph{How does the extrapolation based on EVT compare to the baseline statistical methods?} 

Finally, to establish the usefulness of extreme value theory, it is crucial to study the accuracy of extreme value theory on a set of larger real-world Java web applications. We leverage a set of known side-channel vulnerabilities in critical libraries such as Jetty, Spring Security, and Apache ftp server. In addition, we study how much the prediction of EVT reduces the performance overhead of longer fuzzing campaigns in terms of the number of bytecode executed.

\vspace{0.25 em}
\noindent \textbf{RQ3.} \emph{What are the accuracy and performance gains of EVT for extrapolating the worst-case differential fuzzing on the set of larger Java libraries?}

We find that a combination of the Poisson Process distribution with bootstrapping leads to the best configurations for EVT-based extrapolations. 
Our experiments show that our method outperforms the baseline in 14.3\% cases and achieves competitive results in the remaining 64.2\%. In addition, our approach provides tight upper bounds in 57.1\% of cases, while the competitive baselines are prone to false negatives (underestimation). 
We also find that EVT-based extrapolations do not underestimate the worst-case differences in the fuzzing. The overestimation ranges from 32.8\% to 206.8\%. We show that an EVT-enabled \textsc{DifFuzz} saved the execution of at least a hundred thousand and up to 1 billion bytecode instructions.  

\vspace{0.5em}
\noindent\textbf{Contributions.}
The key contributions of this paper are:
\begin{itemize}[leftmargin=*]
    \item We formalize the connection between differential fuzzing and extreme value theory to extrapolate the maximum differences of an unseen fuzzing campaign;
    \item We infer the parameters and hyperparameters of extreme value theory for Java programs, such as the early stopping criterion for differential fuzzing;
    \item We compare the EVT-based extrapolation to three baseline statistical methods; and
    \item We show the usefulness and performance advantages of EVT-enabled differential fuzzing over larger Java libraries like Spring Security and Apache ftpserver. 
\end{itemize}

%% file: Sections/background.tex
\section{Background and Related Work}
\label{sec:background}
We provide a concise background on the extreme value theory, differential fuzzing, and fuzzing guarantees.

\vspace{0.25 em}
\noindent \textbf{Extreme value theory.} EVT~\cite{coles2001introduction} is a branch of statistics that deals with the analysis of extreme events in a random process.
Given a set of independent and identically distributed random variables $\set{a_1,\ldots,a_n}$, the extreme
value theory is concerned with the min/max statistics of a random process as for instance $M_n = \max(\set{a_1,\ldots,a_n})$ as $n \to \infty$. 

Under some mild assumptions about the smoothness of the distribution via normalizing constants $a_n$ and $b_n$, it has been proved (e.g., see Leadbetter et al.~\cite{leadbetter2012extremes}) that $Pr[(M_n-b_n)/a_n < a] \to G(a)$ as $n \to \infty$ and $G$ belongs to a family of distributions called the \emph{generalized extreme value (GEV)} family.
Each such distribution has the cumulative distribution function (CDF) of the form
\[
G(a) = \exp\left\{-\left[1 + \xi\left(\frac{a - \mu}{\sigma}\right) \right]^{-1/\xi}\right\},
\]
defined over $\{a: 1 + \xi(a - \mu)/\sigma > 0\}$. 
The model has three parameters: a location parameter $-\infty<\mu<+\infty$, a scale parameter
$\sigma > 0$, and a shape parameter $-\infty < \xi < +\infty$.
Depending on the shape parameter $\xi$, the GEV distribution can be classified into three types.
Type I, known as the Gumbel family, defines a subset of GEV distribution when $\xi \to 0$. The tail behavior of type I, $a_{+}$, has infinite support, but the density of GEV decays exponentially (extrapolations are feasible up to a bounded time horizon).
For type II, $\xi > 0$ and $a_{+}$ have infinite support (heavy tail), decaying polynomially, and hence no guarantee \replaced{may}{my} be feasible. Finally, for type III, $\xi < 0$ and $a_{+}$; it has \replaced{bounded set of values}{finite support} (light tail). In this case, the statistical guarantees for the worst-case outcomes are feasible.

\vspace{0.25em}
\noindent \textit{Generalized Pareto Distribution.} 
There are two basic approaches to infer the parameters of GEV distributions; block maximum and threshold approach. The block maximum approach divides samples into blocks of the same size and uses the maximum of each block as the extreme value. Since such an approach is more appropriate for seasonal data, we use the threshold approach, where extreme events exceed some high threshold $u$. In other words, $\{a_i: a_i > u\}$, are extreme values. Labeling these exceedances by $\{d_{(1)},\ldots,d_{(k)}\}$, we define the threshold excesses by $d_j = a_j - u$ for $1 {\leq} j {\leq} k$. It follows that if $Pr[M_n < a] \to G(a)$, then for large enough $u$, the distribution function is approximately:
\[
H(t) = 1 - \left(1 + \frac{{\xi}t}{\hat{\sigma}} \right)^{-1/\xi}
\]
where $t>0$ and $\hat{\sigma} = \sigma + \xi(u - \mu)$~\cite{coles2001introduction}.
This distribution is known as \textit{generalized Pareto} distribution. The implication of the shape parameter $\xi$ is the same as $G(a)$, as a special case of GEV distribution. Our EVT approach follows this distribution.

\vspace{0.25em}
\noindent \textit{Threshold Selection.} 
A proper choice of threshold value $u$ is critical to analyze the
behavior of extreme value distributions. Low values of the threshold $u$ might
include non-tail samples and lead to mixture distributions that violate
the asymptotic basis of the model. On the other hand, high values of the
threshold $u$ might include only a few tail samples and lead to low
confidence in the model. \deleted{due to high variance.} Hence, it is critical
to be confident on the threshold to infer tail distributions.

\vspace{0.25em}
\noindent \textit{Return Levels.}
The inverse of the probability density function of GEV at
probability $p$, is the \textit{return level} $a_p$,  associated with the
\textit{return period} $1/p$. The level $a_p$ is expected
to be exceeded on average once every $1/p$ period of time.
A \textit{return level} is represented with {$(m,a_m)$}, where $m$ is the time period (E.g. the number of iterations in a fuzzing campaign) and the level $a_m$ is the expected extreme value during the $m$ period (e.g., expected worst-case differences in the next $m$ interactions \deleted{of a fuzzing campaign}).

\vspace{0.25em}
\noindent \textbf{Differential Fuzzing.}
%
Differential fuzzing for side-channel vulnerabilities with \dfuzz~\cite{8812124} deploy the concept of self-composition~\cite{barthe2011} to identify violations of the non-interference principle. Given a side channel such as execution time, \dfuzz attempts to find two secret inputs $z_1$ and $z_2$ and one public input $x$ so that the executions of the program under test $\Pp$ lead to different observations along the side channel $c$: $c(\Pp[x,z_1]) \neq c(\Pp[x,z_2])$. Since the only difference between the two executions is the secret input, an observed difference in the execution behavior along a particular side channel must originate from a secret-dependent path. This does not guarantee that a vulnerability can be exploited, but indicates a vulnerability that should be further investigated. To find inputs that expose this behavior, \dfuzz uses a custom graybox fuzzing algorithm that is guided to maximize the cost difference $\delta = |c(\Pp[x, z_1]) - c(\Pp[x,z_2])|$ between executions.

\noindent \textbf{Statistical Guarantees for Fuzzing.} 
\added{While there are significant efforts (e.g., ReFuzz~\cite{electronics10161921} to improve the efficacy of graybox fuzzing, we focus on fuzzing techniques with statistical guarantees on stopping criteria.}
In the context of black-box fuzzing, Woo et al.~\cite{bb_sched} model the fuzzing process as a weighted coupon collectors problem~\cite{motwani1995randomized}. A fuzzer discovering a new path or revisiting the same path is akin to collecting seen or unseen coupons. The STADS framework~\cite{böhme2018stadssoftwaretestingspecies} draws similarities between the ecological question of estimating the number of rare species in an area to the number of rare program paths that may contain unseen bugs. Hence, a practitioner can use the STADS to estimate rare paths based on discovered ones during a graybox fuzzing campaign. Residual risk estimation \cite{residual_risk} extends the STADS framework to quantify the risk of stopping the fuzzing campaign early and the probability of discovering the next unseen path. Similarly, estimation saturation in fuzzing \added{(Reachable Coverage~\cite{reachable_cov})} provides a statistical approach to the problem of estimating reachable paths as a stopping criterion. \added{Green Fuzzing~\cite{10.1145/3597926.3598043} uses the coverage of potentially vulnerable code fragments rather than the entire code basis as a saturation-based stopping criterion.}
Rare path guided fuzzing~\cite{saha2023rare,10.1145/3510003.3510227} uses a combination of probabilistic analysis and symbolic methods to identify rare paths to guide graybox fuzzing. \replaced{While these works estimate the maximum trials to cover all reachable paths in the control-flow graphs, our work aims to find the maximum differences between the execution of two similar inputs (i.e., the number of trials to observe the maximum cost differences between two executions of similar inputs. Furthermore, information-flow graphs are critical to our stopping criteria of differential fuzzing, but none of these prior works integrate information flow to estimate when to stop fuzzing. 
An example is the following program: \texttt{void apply\_and\_wait(bit[32] secret, bit[32] guess): return sleep(secret \& guess)} where there is only one control-flow graph path, but there are $65,536$ different waits (costs). The maximum cost difference between two executions is $2^{16}$, and it happens when one run has been executed with secret=0...0 and guess=1...1, while another has been executed with secret=1...1 and guess=1...1. Our approach uses EVT to extrapolate the maximum cost differences from observations during differential fuzzing.
}{To the best of our knowledge, no prior work has presented an estimate of risk in differential fuzzing.} 

EVT has been widely used to provide probabilistic guarantees on worst-case execution times in real-time and embedded systems~\cite{lu2011new,cucu2012measurement,hansen2009statistical,10.1145/3644815.3644989}. Prior work has explored EVT's application to timing analysis with random caches, studied the impact of relaxing i.i.d. assumptions, and developed techniques for handling measurement challenges on modern hardware. EVT has been used to detect rare bugs in circuit design\cite{singhee2007statistical} and to estimate the worst-case delay of VLSI circuits~\cite{8342220}.
While these studies demonstrated EVT's effectiveness for timing analysis in embedded systems, they focused primarily on hardware-level timing behaviors and single worst-case execution, rather than software testing applications.\deleted{like differential fuzzing.}

%% file: Sections/overview.tex
\section{Overview}
\label{sec:overview}

Let us first consider a code snippet taken from a vulnerable password matching program in Apache WSS4J~\cite{WSS4J}.
Figure~\ref{fig:jetty-streql} shows the code where the secret string $s$ is compared to a given public guess string $g$ that has the same length (32 characters are used for this example). We also abstract the costs with the number of bytecode instructions executed.

\input{Sections/overview-code}

\begin{figure*}[tbp!]
    \centering
    \begin{subfigure}[b]{0.32\textwidth}
        \includegraphics[width=\textwidth]{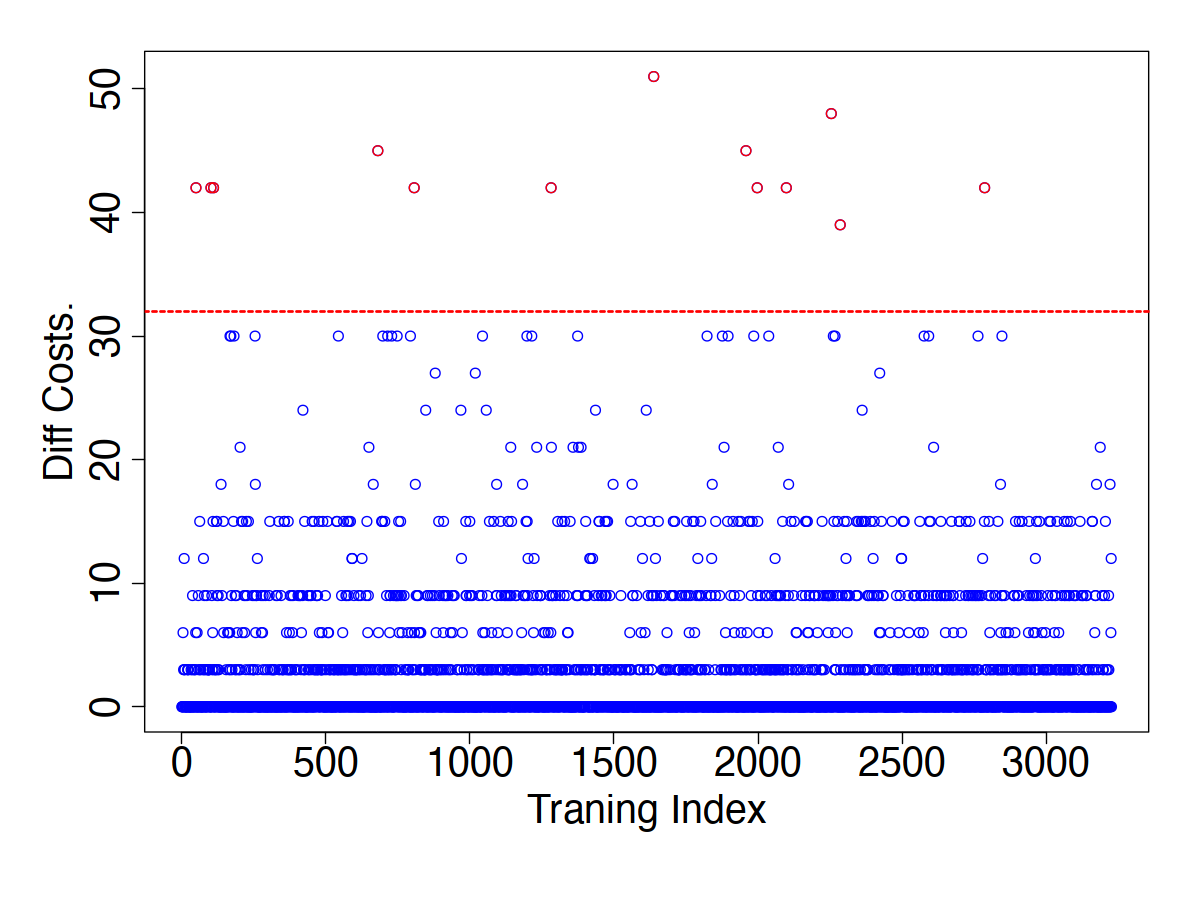}
        \caption{Training samples and the threshold.}
        \label{fig:SB_threshold}
    \end{subfigure}
    \begin{subfigure}[b]{0.33\textwidth}
        \includegraphics[width=\textwidth]{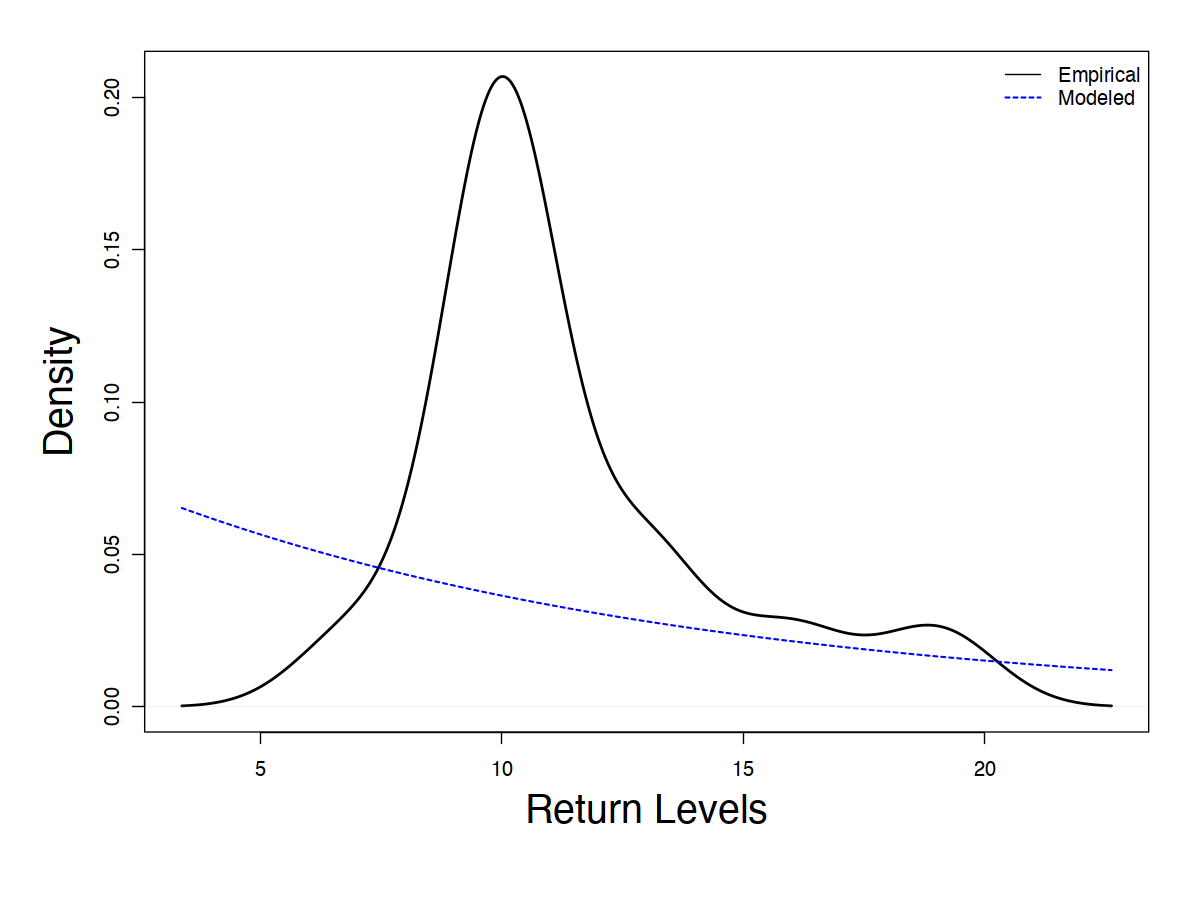}
        \caption{Tail Distribution with Exponential Basis.}
        \label{fig:SB_return_level}
    \end{subfigure}
    \begin{subfigure}[b]{0.33\textwidth}
        \includegraphics[width=\textwidth]{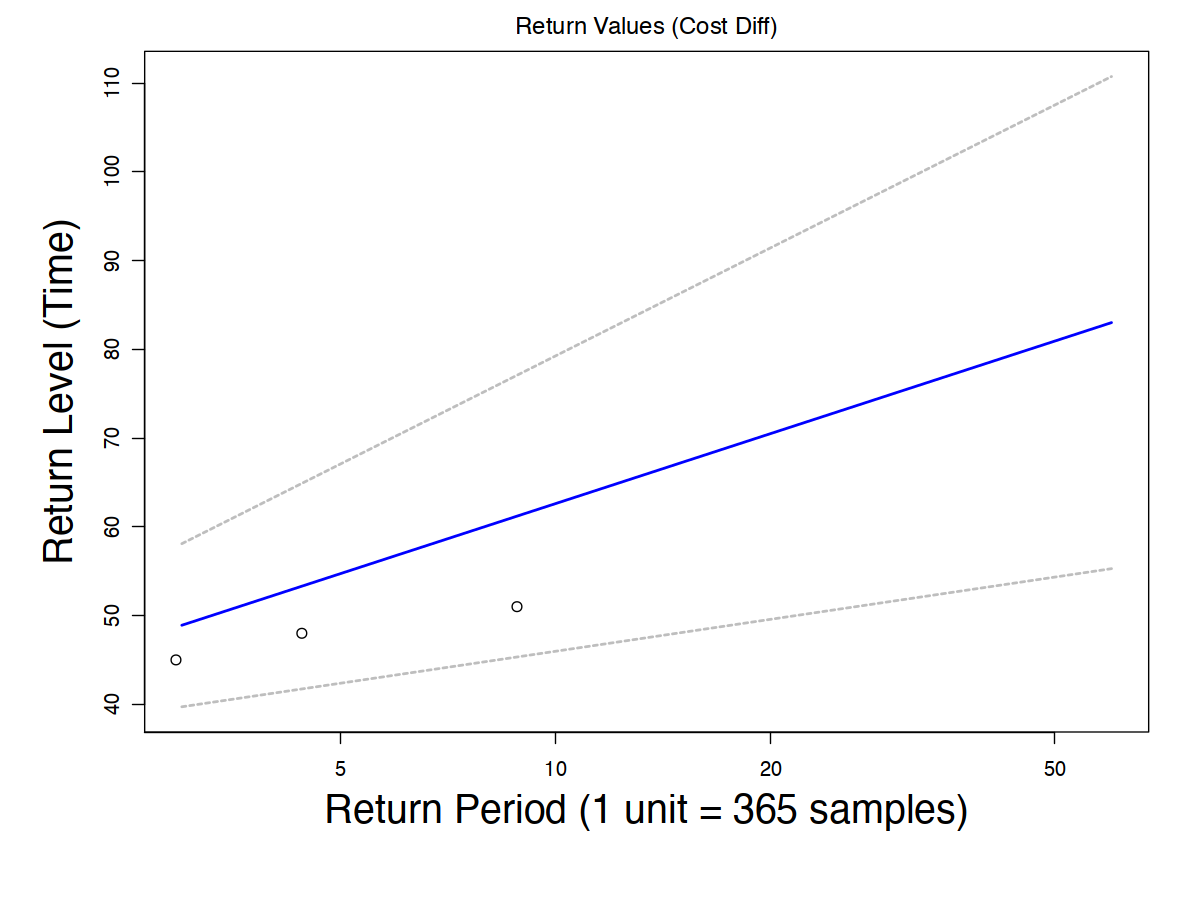}
        \caption{Return Levels of Max. Cost Differences.}
        \label{fig:SB_density}
    \end{subfigure}

    \caption{Overview Example. (a) The cost differences over training samples (the first 3,226 samples in \dfuzz).
    (b) The empirical tail distribution of \dfuzz. 
    (c) m-return level plot of cost differences with expected values (and their 95\% CI).}
    \label{fig:overview}
\end{figure*}

In this paper, we leverage \dfuzz for our analysis.
We run the fuzzer for $1$ hour and collect 23,226 samples. \dfuzz satisfies exponentiality testing at the iteration 3,226. We ran fuzzing further for 20,000 more iterations to collect the ground truth. In other words, we used the first 3,226 samples as training samples, and the rest of fuzzing as testing data samples. 
The maximum cost observed over the training samples is 69. The mean and standard deviations over the training samples is 2.9 and 5.6, respectively. We use the training samples to extract GEV distributions. In doing so, the first step is to find a suitable threshold value such that any values that exceed this threshold belong to the tail distribution of differential fuzzing.
We used a bootstrapping technique (see details in Approach section~\ref{sec:approach}) that takes the training samples and infers a suitable threshold. For the example of Apache WSS4J, the threshold sets to 32, meaning that any bytecode differences above 32 are deemed extreme values. Figure~\ref{fig:overview} (a) shows the cost differences in training set, the threshold for extreme values, and samples that belong to tail distributions (13 samples).

\begin{table}
\centering
 \caption{Return Levels of max. costs of graybox fuzzing (WSS4J password matching).}
\resizebox{0.49\textwidth}{!}{
 \begin{tabular}{ | c | c  c  c| } 
  \hline
 \textbf{Level} & Return Level (\textbf{observed}) & Return Level (\textbf{Exp.}) & Error of Exp. (\%) \\ 
 \hline 
 1,000 & 84 & 43.3 [36.3, 50.3] & -40.1  \\
 \hline
 2,000 & 84 & 50.3 [38.9, 61.6] & -26.6 \\
 \hline
 5,000 & 96 & 59.4 [42.4, 76.4] & -20.4  \\
 \hline
 10,000 & 96 & 66.3 [45.1, 87.6] & -0.09 \\
 \hline
 20,000 & 96 & 73.3 [47.7, 98.9]  & +0.03 \\
 \hline
  \end{tabular}
 }
 \label{tab-RL-IID-password-256}
 \label{tab-RL-GB-password-256}
\end{table}

Once the threshold value is inferred, we choose a type of extreme value distribution and infer its parameters.
For this benchmark, we use Exponential and Poisson Process (PP) types and infer their parameters via Bayesian optimization.
Figure~\ref{fig:overview} (b) shows the empirical tail distribution with exponential distribution. The location and shape of distributions are 36.4 (+/- 2.4) and 11.4 (+/- 3.2), respectively. One crucial aspect of GEV analysis is the concept of ``return-level.'' It shows the expected maximum value for a given time period in the future. We use the concept of return levels to extrapolate the expected maximum cost differences in the next $m$-iterations of \dfuzz. Figure~\ref{fig:overview} (c) shows the return levels where one unit is 365---prevalent EVT models use the number of days in a year---fuzzing iterations. 
Table~\ref{tab-RL-IID-password-256} shows the results of GEV-based extrapolations (\textbf{Prediction}) at  iteration 3,226 of fuzzing, when the exponentiality testing passed. In the next 20,000 fuzzing iterations, we observe the maximum cost difference of 96 bytecodes, whereas the GEV prediction shows a maximum cost difference of 73.3 [47.7, 98.9], which is an overestimation of 0. 03\%. 
Overall, as the number of fuzzing iterations increases, the prediction provides better estimates.

\begin{figure*}[tbp!]
    \centering
    \begin{subfigure}[b]{0.48\textwidth}
        \includegraphics[width=\textwidth]{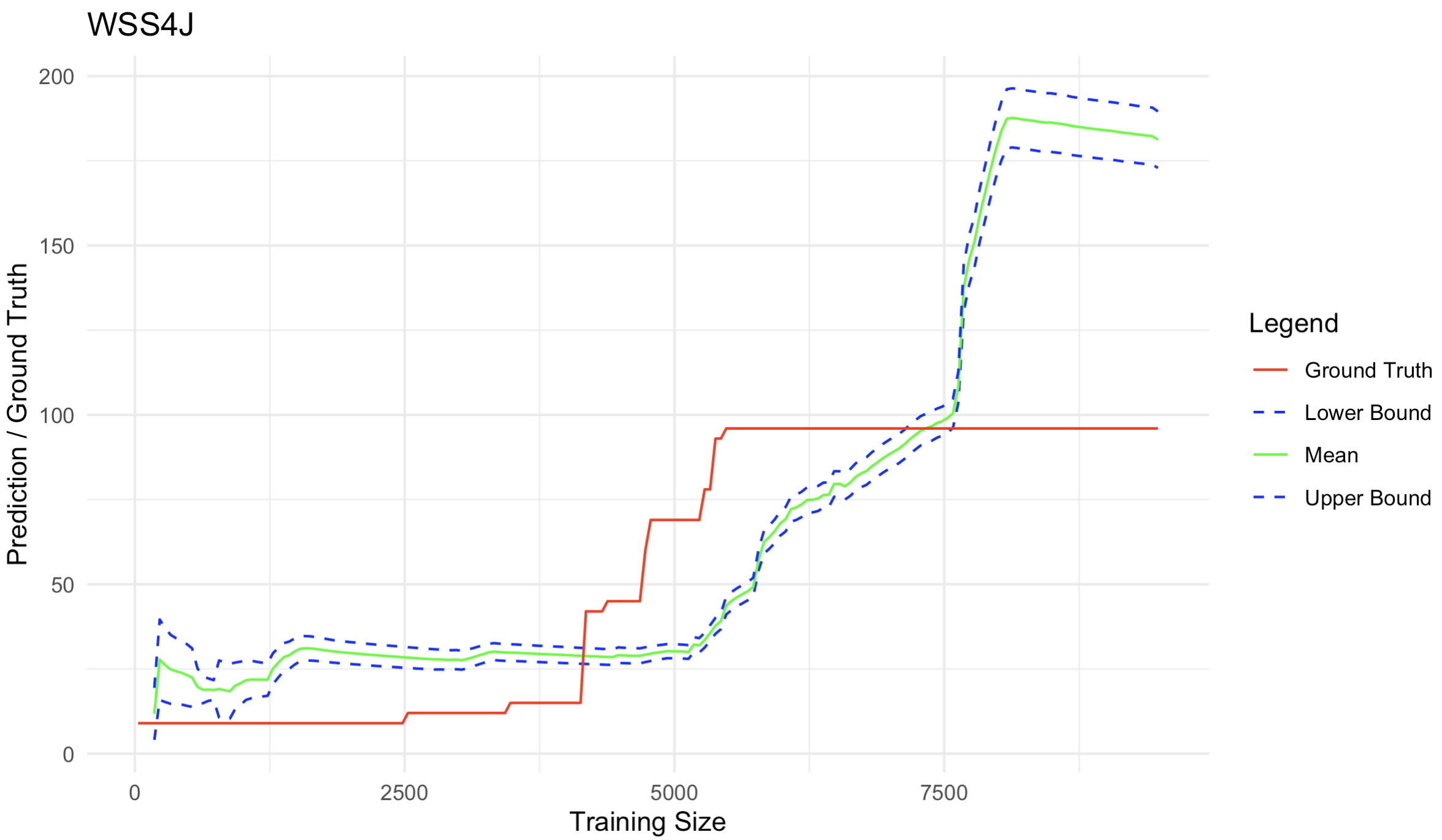}
        \caption{Predicting via Exponential Basis.}
        \label{fig:WSS4J_Exp}
    \end{subfigure}
    \begin{subfigure}[b]{0.48\textwidth}
        \includegraphics[width=\textwidth]{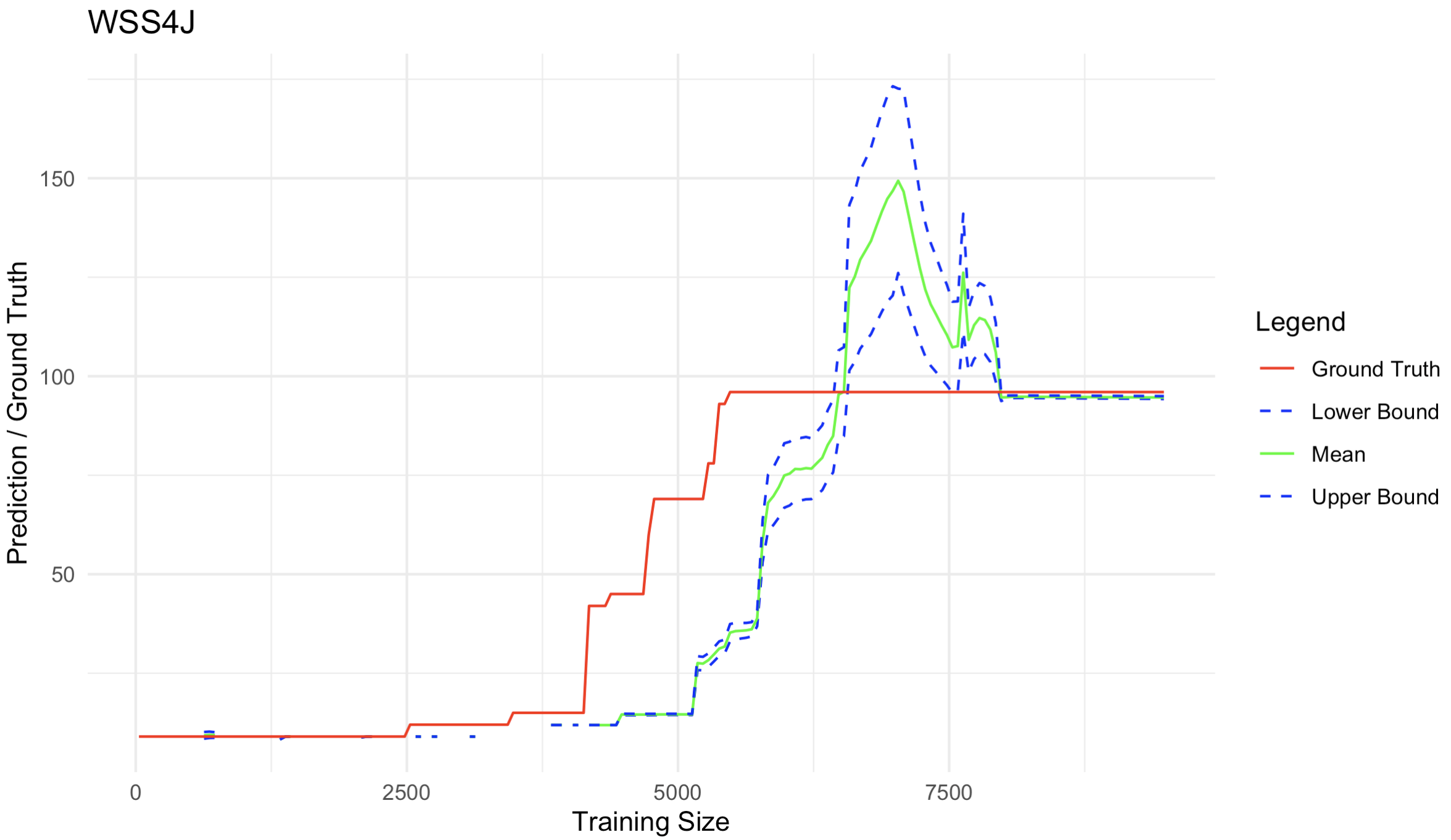}
        \caption{Predicting via Poisson Process.}
        \label{fig:WSS4J_PP}
    \end{subfigure}
    \hspace{0.5em}
    \caption{Temporal Plot of Prediction. We use the size of training (x-axis) to predict the max difference in the next 1,000 fuzzing iterations (green) as compared to the ground truth (red) with Exponential and PP distributions.}
    \label{fig:WSS4J_temporal}
\end{figure*}

Figure~\ref{fig:WSS4J_temporal} shows the temporal progress of maximum cost differences vs.~predictions with EVT with exponential (Figure~\ref{fig:WSS4J_Exp}) and PP (Figure~\ref{fig:WSS4J_PP}) bases for the first 10,000 fuzzing iterations.
Each point in the x-axis shows the number of iterations used to infer EVT distributions, and the corresponding value in the y-axis shows the 95\% confidence intervals of the prediction (green series with blue intervals) vs. the ground truth max. differences (red series) up to next 1,000 iterations. The plots show that the PP is more sensitive to changes in the distribution of cost differences and can predict the ground truth more accurately. On the other hand, exponential is more conservative and may provide a sound upper-bound. \deleted{on the ground truth.}

%% file: Sections/overview-code.tex
\begin{figure}[t!]
\centering
\begin{lrbox}{\mybox}%
\begin{scriptsize}
\begin{mylisting}[hbox,enhanced,drop shadow]{stringEquals (WSS4J)}
boolean ?{\textbf{stringEquals}}?(String s1, Object s2)
{        
?\indentrule? if (s1 == s2) {
    return true;
?\indentrule? }
?\indentrule? if (s2 instanceof String) {
    String s2Str = (String)s2;
    int n = s1.length();
    if (n == s2Str.length()) {
        char v1[] = s1.toCharArray();
        char v2[] = s2Str.toCharArray();
        int i = 0;
        while (n-- != 0) {
            if (v1[i] != v2[i])
                return false;
            i++;
        }
        return true;
    }
?\indentrule? }
?\indentrule? return false;
}
\end{mylisting}
\end{scriptsize}
\end{lrbox}%
\scalebox{0.85}{\usebox{\mybox}}

\caption{String equality in Apache WSS4J ({\tt s1} secret, {\tt s2} public).
The code snippet is the implementation for the secret comparison with a given public guess.
}
\label{fig:jetty-streql}
\end{figure}

%% file: Sections/problem-statement.tex
\section{Problem Statement}
\label{sec:problem}
We consider differential testing methods that take a program $\Pp$ and their input domain variables $\Aa$ and search to find two inputs that share a common property $\phi$, but their executions on the program $\Pp$ lead to a maximum difference.
For example, a side-channel fuzzer like \texttt{DifFuzz}~\cite{8812124} takes a program $\Pp$ with two set of input variables, a secret set of variables $Z$ (e.g., stored secret password) and a public set of variables $X$ (e.g., a guess for a password) and tries to find two inputs with the common public input values which led to a maximum difference in the side-channel observation (e.g., timing) of program $\Pp$ due to a difference in the secret input, i.e., \texttt{DifFuzz}($\Pp$, time, $Z$, $X$) ::= $\max_{z_1,z_2,x}. |\Pp_{time}(x,z_1)~{-}~\Pp_{time}(x,z_2)|$, hence it shows the presence of (strong) leaks of secrets via side channels.

One critical limitation of differential testing and fuzzing techniques, due to their dynamic execution nature, is that they are prone to false negatives and cannot provide guarantees on the worst-case divergence. 
Let $\Delta$ = $\{\delta_1,\ldots,\delta_m\}$ be the set of random variables following an underlying unknown distribution of differentials, where $\delta_i = |\Pp(a)~{-}~\Pp(a')|$ shows the difference at step $i$ of fuzzing. Practically, we try fuzzing for at most $m {<<} n$ iterations and are interested in estimating \texttt{max}($\delta_{m+1}$, $\ldots$, $\delta_n$), assuming that our fuzzing process has become stationary at step $m$ of fuzzing. 
Following the central limit theorem as $n \to \infty$, the sum of random variables follows a Gaussian distribution, i.e., $\mathcal{N}(\mu, \sigma)$ where $\mu = \texttt{average}(\Delta)$ and $\sigma = \texttt{std}(\Delta)$. However, infamous concentration results (e.g., Markov and Chebyshev Inequalities ~\cite{motwani1996randomized}) provide tail guarantees on the \textit{expected} cost differences; while we are interested in the statistical guarantees on the \textit{maximum} differences.  

\vspace{0.5 em}
\noindent \textbf{Statistics of Tail Distributions.}
Rather than modeling the distributions of expectations, we are interested in the tail distribution, i.e., $M_n = \max(\set{\delta_1,\ldots,\delta_n})$.
In fact, the testing and fuzzing campaigns are exploring the tail distribution of differentials without any explicit model of such distributions. In the same way that the central limit theorem relates the sampling distribution of expectation to Gaussian distribution; EVT connects the sampling distribution of maximum of random variables to the GEV family of distributions.

\begin{definition}[Tail Distributions of Differential Fuzzing]
    \label{def-problem}
     Given a differential fuzzing technique that takes a program and searches its space to find inputs that characterize the maximum differences between similar inputs after $m$ iterations (e.g., after satisfying a statistical testing or after a time-out), our goal is to model the tail distribution of differential fuzzing and extrapolate ``what would be the (expected) worst-case difference $\delta_n$ if the fuzzer had run for $n$ more iterations?''   
\end{definition}

%% file: Sections/approach.tex
\section{Approach}
\label{sec:approach}
Our key approach is to leverage EVT to model the tail of differential fuzzing processes. The key advantage is that EVT directly models the tail distribution\deleted{(as opposed to modeling the tail of regular distributions like Gaussian distribution)}, allowing us to reason about the validity and extrapolate the return levels $\delta_n$ of extreme values for a time period of $n$.

\begin{algorithm}[t!]
\DontPrintSemicolon
    \KwIn{Program $\Pp$, Fuzzing infrastructure \textsc{Fuzz}, 
    Statistical Test of Tail \texttt{STT},
    Worst-Case Differential Predictions \texttt{WCDiff},
    Time-out $\mathcal{T}$.
    }
    
    $\Delta$, $Pred$, $i$, $t$ $\gets$ $\{\}$, -1, 0, \texttt{current}(time)
    
    \While{\texttt{current}(time) $\leq t+\mathcal{T}$}{
        $z_1, z_2, x$ $\gets$ \textsc{Fuzz}($\Pp$, $max_{inp}$($\Delta$))

        $cost_1$ $\gets$ \textsc{measure}($\Pp$, $z_1$, $x$)

        $cost_2$ $\gets$ \textsc{measure}($\Pp$, $z_2$, $x$)

        $\delta$ $\gets$ $|cost_1 - cost_2|$

        $\Delta$.\texttt{add}($\delta$, [$x,z_1,z_2$])

        \If{\texttt{STT}($\Delta$)}{

            $Pred$ $\gets$ \texttt{WCDiff}($\Delta$)

            \Return  $\Delta$, $Pred$, $i$

        }
        $i$ $\gets$ $i$ + 1
    }

    \Return  $\Delta$, $Pred$, $\mathcal{T}$
\caption{EVT-enabled \dfuzz algorithm.}
\label{alg:overall}
\end{algorithm}

Algorithm ~\ref{alg:overall} shows the overall approach that is an extension to differential fuzzing \dfuzz~\cite{8812124} with statistical guarantees. It takes a target program $\Pp$, a basic fuzzing \textsc{Fuzz}, a statistical testing of tail \texttt{STT}, a worst-case predictor \texttt{WCDiff}, and the maximum allowed time or number of iterations for the analysis $\mathcal{T}$ as inputs. In the first step, the algorithm begins by initializing $\Delta$ as an empty set to store differences and $Pred$ to -1 to extrapolate the worst-case differences. 

Then, a \texttt{while} loop is started to run as long as the current time $t$ does not exceed the time-out limit $\mathcal{T}$. 
In each iteration, it uses the differential fuzzing tool \dfuzz to generate two similar inputs $(x,z_1)$
and $(x,z_2)$ that only differ in some sensitive features
by querying the program $\Pp$. Then, it measures the cost of each execution (I.e., in terms of executed bytecodes), and stores the results in $cost_1$ and $cost_2$.
Then it calculates the absolute difference $\delta$ between the two measured costs. \dfuzz finally adds the computed difference to the set of differences $\Delta$ as the feedback to the fuzzing engine besides other metrics such as the input visited a new path in the control flow graph. 

\begin{figure*}
    \centering
\includegraphics[width=0.85\linewidth]{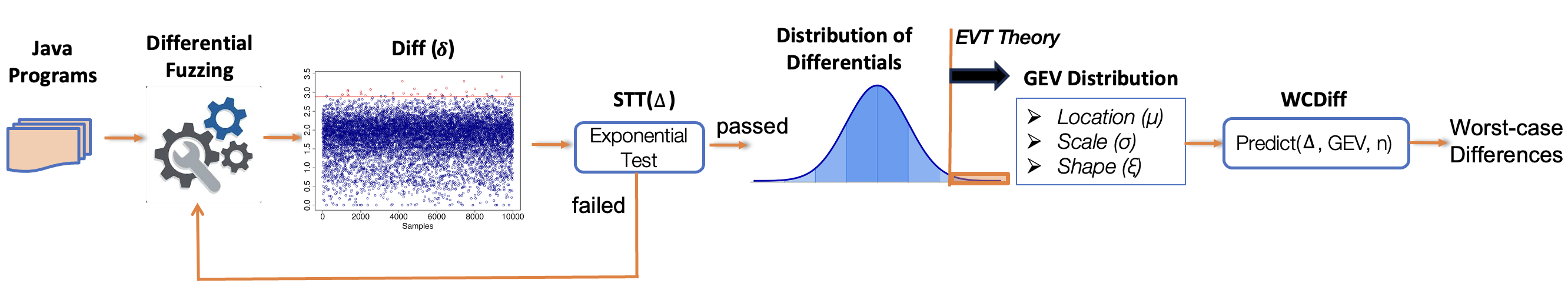}
    \caption{\added{The Conceptual Diagram of Algorithm~\ref{alg:overall}: Steps to infer the worst-case differences.}}
    \label{fig:overview-diagram}
\end{figure*}

Next, our approach performs a statistical testing (\texttt{STT}) during the fuzzing, and if the collected samples satisfy the test, we query the prediction models (\texttt{WCDiff}) that model the tail distributions of cost differences to extrapolate the maximum differences in a given number of fuzzing trials, and we terminate the fuzzer. To collect the ground truth differences, in practice, we continue fuzzing to record the differentials up to the time-out. \noindent \added{ Figure~\ref{fig:overview-diagram} visualizes Algorithm~\ref{alg:overall} to infer the worst-case difference in a differential fuzzing via EVT.}

\subsection{Statistical Testing of Tail (\texttt{STT})}
There are multiple statistical tests on the tails of random processes to convince a reliable distribution of tails~\cite {doksum198426}.
We consider two tests: 1) Laplace and 2) Exponentiality.

\vspace{0.5 em}
\noindent \textbf{Laplace.} The widely recognized Laplace estimator assigns a small probability to unobserved events by treating each one as if it had been observed exactly once. Pierre-Simon Laplace applied this approach to tackle the sunrise problem; given that the sun has risen for \(n\) consecutive days up to today, what is the likelihood it will rise again tomorrow? This line of inquiry led Laplace to formulate the rule of succession, laying the essential groundwork for Bayesian statistics.

Let $\delta_1,\ldots,\delta_n$ be a sequence of cost differences observed during fuzzing. Let $i$ be the index of a random variable that exceeds any previous cost differences, i.e., $\delta_i > \max \set{\delta_1,\delta_{i-1}}$. Following the Laplace estimator, with a probability $p=\frac{1}{j+1}$, we can stop fuzzing if and only if $\max_{i} \set{\delta_{i+1},\delta_{i+j}} \leq \delta_i$. One can set $j$ to $100$ to bound the max cost difference with at most, 0.05 probability.  

\begin{algorithm}[t!]
\DontPrintSemicolon
    \KwIn{Differentials $\Delta$, Min. Number of Samples $k_{min}$, Max. Number of Samples $k_{max}$.}
    
    \If{\texttt{size}($\Delta$) $<$ $k_{max}$}{
        \Return  \texttt{False}
    }

    $res$ $\gets$ \texttt{True}

    \For{$k$ $\gets$ $k_{min}$ to $k_{max}$}
    {

        $\Theta$ $\gets$ \texttt{Select\_Top\_k}($\Delta$, $k$)

        $\overline{\Theta}$, $\sigma(\Theta)$  $\gets$ \texttt{average}($\Theta$), \texttt{std}($\Theta$)

         $CV_k$ $\gets$ $\frac{\sigma(\Theta)}{\overline{\Theta}}$
         
         \If{$CV_k$ $\geq$ $1.0 + (\frac{1}{4*k})$}{
         
            $res$ $\gets$ \texttt{False}                
         
            \textbf{break}
         }
    }

    \Return  $res$
    
\caption{\texttt{Exponentiality} as a type of \texttt{STT}}
\label{alg:exp}
\end{algorithm}

\vspace{0.5 em}
\noindent \textbf{Exponentiality.} This test utilized the Coefficient of Variation (CV) to determine whether the tail distribution is well-behaved. Algorithm ~\ref{alg:exp} shows the steps in performing exponentiality testing. Specifically, the test goes over the $k$ highest values of the cost differences and calculates the CV value where $k$ ranges from $k_{min}$ to $k_{max}$. If for all values of $k \in [k_{min},k_{max}]$, the CV is less than
($1.0 + \frac{1}{4*k}$), then we are statistically confident that we have enough samples from the tail to infer a well-behaved tail distribution~\cite{castillo2014methods}. Note that the extra term ($\frac{1}{4*k}$) is to correct the bias in the estimation of CV due to small sample size in the tail~\cite{sokal1995biometry}. 
Otherwise, if any values of CV are greater than ($1.0 + \frac{1}{4*k}$), we may not be able to infer a well-behaved distribution in the tail. Hence, we return \texttt{False}, it requires further fuzzing iterations. 

\begin{algorithm}[t!]
\DontPrintSemicolon
    \KwIn{Differentials $\Delta$, Threshold Finding Method $T$, Type of EVT Distribution $\Dd$}

    Threshold $\gets$ 0.0
    
    \If{$T$ == `BootStrap'}{
        $t$ $\gets$ \texttt{quantile}($\Delta$, seq(0.99, 0.75, by = -0.01))

        $\Delta_{bs}$ $\gets$ $\lambda$. $t$ \texttt{sample}($\Delta[\Delta > t]$, R=\texttt{True}, 1000)

        $GPD$ $\gets$ $\lambda$. $\Delta_{bs}$ \texttt{fit}(`gpd', $\Delta_{bs}$)
            
        Threshold $\gets$ $\lambda$. $t,\Delta_{bs}$ \texttt{min}$_t$($GPD$.parameters.CI($\Delta_{bs}$)) 
    }
    
    $GEV$ $\gets$ \texttt{fevd}($dist$ = $\Dd$, $pot$ = Threshold)

    location, scale, shape $\gets$ \texttt{distill}($GEV$)

    \If{shape is valid}{
        Prediction $\gets$ $\lambda$.period \texttt{ReturnLevel}(location, scale, shape, period)
    }\Else{
        Prediction $\gets$ $\max(\Delta)$
    }

    \Return Prediction

\caption{Prediction of worst-case differentials via a generalized extreme value distribution.}
\label{alg:gev}
\end{algorithm}

\subsection{Extrapolations via Tail Distributions (\texttt{WCDiff})}
Once a statistical testing of tail is convinced, the next step is to infer the tail distribution and estimate the return levels of worst-case cost differences.

\vspace{0.5 em}
\noindent \textbf{Bayes Factor.} Following the standard hypothesis testing,
one can come up with two hypotheses where the null hypothesis is a predicate that the cost differences are below a threshold and the alternative hypothesis is the negation of such predicate. The null and alternative hypotheses are
\[
\Hh_0: \pP(\delta) \leq \tau,~~\Hh_1: \pP(\delta) >  \tau
\]
where $\pP(\delta)$ is the probability that a cost difference $\delta$
stays below a threshold $\tau$, $\Hh_0$ is the null hypothesis, and $\Hh_1$ is the alternative. We say a Bayes factor has passed if we witness enough 
samples that are below the threshold $\leq$ to accept $\Hh_0$
as opposed to $\Hh_1$. There are multiple ways to conduct such statistical testing. 
The sequential probability ratio test and a Bayes factor are examples. We follow Jeffreys test~\cite{jha2009bayesian,sankaranarayanan2013static}, a variant of Bayes factor, with a uniform prior to find a lower-bound on the number of successive samples $K$ that sufficient for us to convince $\Hh_0$:
\[
K \geq \lceil(-\log_2 B)/(\log_2 \theta)\rceil  
\]
where $B$ is Bayes factor and can be set to $100$ for very strong evidence. For instance, to achieve $\theta = 0.95$, we need to set $K \geq 90$ to accept $\Hh_0$.

\vspace{0.5 em}
\noindent \textbf{Concentration Inequalities.}
One natural idea is to leverage concentration inequalities such as Markov's Inequality, Chebyshev’s Inequality, and H$\ddot{o}$ffding Bound~\cite{motwani1995randomized}. 
Markov's Inequality states that for a non-negative random variable $\delta$, the probability that it exceeds the expectation $\mu$ by a factor of $k$ times are less than $\frac{1}{k}$, i.e., 
$Pr[\delta > k.\mu] \leq \frac{1}{k}$.

Similarly, Chebyshev’s Inequality states that for a non-negative random variable $\delta$ with an expectation $\mu$ and variance $\sigma^2$ and for any real number $k > 0$, we have ${\displaystyle \Pr(|\delta-\mu |\geq k\sigma )\leq {\frac {1}{k^{2}}}.}$
One critical limitation of these inequalities in our setting is the focus on the expectation and how much a random variable can exceed the expectation. While we are interested in modeling and reasoning about the tail of random variables; these concentration
results do not explicitly model the tail distribution. 

Rather than analyzing the concentration of differentials, we propose to explore the tail distribution.  Algorithm~\ref{alg:gev} aims to predict worst-case differentials using EVT. 
The algorithm processes a dataset of observed differentials $\Delta$ to model the extreme values and estimate the maximum expected differential over specified periods. 
The algorithm also takes the method for selecting threshold $T$ and the type of tail distribution $\Dd$ as inputs.
If the threshold inferring method is ``BootStrap'', it proceeds with the bootstrap method to determine an optimal threshold value (any values above the threshold are extreme). 
The ``BootStrap'' generates a sequence of quantile thresholds $t$ ranging from the 99th percentile to the 75th percentile of $\Delta$, decreasing by 1\% increments. This creates multiple candidate thresholds for EVT modeling.
For each threshold $t$, it performs bootstrap sampling on the exceedances (data points where $\Delta [\Delta > t]$) and samples $1000$ data points with replacement (bootstrap sampling) from the exceedances.
Then, for each set of exceedances (based on the threshold $t$), we fit Generalized Pareto Distribution (GPD) over the samples and infer the parameters of distributions and their confidence intervals.
Following the BootStrap method, if the parameter estimations for GPD are tight (a narrow confidence interval for a valid distribution), then it considers the corresponding threshold $t$ as the optimal threshold that provides the best statistical fit. 
Finally, the algorithm fits a generalized extreme value (GEV) distributions to the data using a type of distribution ($\Dd$=Poisson Process, exponential, etc.) with the optimal threshold $t$ as the peaks over threshold value.
The algorithm extracts the parameters of GEV and validates its behavior based on its shape. In particular, if the shape is zero or negative ($\xi <= 0$), then the GEV belongs to the type I (exponential) or type III (light), and an extrapolation is feasible. Finally, the algorithm calculates the return levels of extreme values based on the parameters of GEV distributions for a given return period (i.e., the number iterations in fuzzing). The outcome is the prediction for the worst-case cost differences.

%% file: Sections/experiments.tex
\section{Experiments}
\label{sec:experiment}




\subsection{Research Questions.}
In this paper, we study the following research questions:
\begin{enumerate}[start=1,label={\bfseries RQ\arabic*},leftmargin=3em] 

\item What are the best statistical testing to stop fuzzing to predict the worst-case cost differences and infer a tail distribution of fuzzing process with ideal configurations of extreme value distributions? 

\item Do GEV distributions predict the worst-case cost differences better than the baseline such as Markov's Inequality~\cite{chernoff1952measure} and  Chebyshev’s Inequality~\cite{chebyshev1867valeurs}?

\item How accurate is the EVT in predicting the maximum differences on larger (real-world) Java libraries, and what are the efficiency characteristics? 

\end{enumerate}

\subsection{Subjects}
\label{sec:subjects}
For RQ1, we use a set of micro-benchmark to infer ideal statistical techniques and their hyperparameters.
In particular, we compare two statistical testing methods as stopping criterion during fuzzing,
two threshold finding methods, and two types of EVT distributions over \textit{Leak Set} programs that are leaking the number of set bits. In \textit{Leak Set}, the size of secrets are from $12$ to $28$ bits for \textit{Leak Set 1} to \textit{Leak Set 5}, respectively. In RQ2, we compare our EVT-based extrapolations to three primary statistical methods.
We used benchmarks from  \blazer~\cite{antonopoulos2017decomposition}, \themis~\cite{DBLP:conf/ccs/ChenFD17}, and \diffuzz~\cite{8812124} that include vulnerabilities in \textit{Eclipse Jetty} and \textit{Apache WSS4J}\footnote{\url{https://issues.apache.org/jira/browse/WSS-677}}.
Finally, RQ3 includes larger-scale Java libraries such as \textit{Spring Security} and \textit{Apache Ftpserver} 
where we used the best statistical techniques and their
configurations to investigate the accuracy of our approach in predicting the worst-case differential
fuzzing and evaluate the performance gain due to early stopping of fuzzing. 

\subsection{Technical Details}
Experiments were ran on an Amazon AWS EC2 m5.large instance with Ubuntu $18.04.1\,LTS$
featuring $2x$ Intel(R) Xeon(R) CPU $X5365$ @ $3.00GHz$ with $8GB$ of memory, \textsc{OpenJDK} $1.8.0$\_$422$ and GCC $9.4.0$.
Following the setup for \dfuzz~\cite{8812124}, we run each benchmark five times and report the mean and standard deviations of the results. We also run each of the experiments for 30 minutes. We apply Mann–Whitney U test~\cite{mann1947test} to establish that an error of approach 1 is statistically less than another and vice-versa.

\input{Sections/Table_RQ1}
\subsection{RQ1 -- Inferring ideal configurations for EVT prediction}
\label{sec:experiments:RQ1}
Our goal is to infer which statistical testing during fuzzing provides an ideal sample set to infer the parameters of extreme value distributions. We modify \diffuzz to implement two methods of early stop: 1) Laplace and 2) Exponentiality. 
Our evaluation quantifies the error in the estimation of worst-case differential costs as compared to the ground truth.
We note that while we set the training sample size to the time when the statistical testing for early stopping is satisfied,
we continue fuzzing to record the ground-truth for our evaluations. Then, we use two techniques to infer thresholds of GEV distributions: 1) bootstrapping and 2) 0.95-Quantile. Finally, we consider two types of GEV distributions: 1) Exponential (Exp.) that represent the tail when shape is zero (i.e., it assumes a infinite tail, but decaying), and 2) Poisson Process (PP) that is valid for non-positive values of shape (i.e., the shape is finite). Therefore, we exclude invalid results that do not satisfy these invariants. 

Table~\ref{table:RQ1} shows the performance of different methods. We report the total number of test cases generated in 30 mins fuzzing campaign, the training (when the statistical testing passed during fuzzing) vs. testing (when we continue fuzzing after meeting the early stop criterion to collect the ground truth), the ground truth maximum cost differences, and the EVT-based prediction. We also report the most accurate predictions (over the repeated benchmarks) as well as the least error.   
We divide the table into different parts and use the least error to identify superior techniques. 
In doing so, we highlight any predictions that are within 5\% of ground truth. Since we prefer over-approximation of ground-truth over the under-approximation, we also highlight any results that over-approximate the ground truth by at most 10\%.
For example, in \textit{Leak Set (1)}, with Exponentiality testing, bootstrapping threshold, and PP distribution type; the best EVT extrapolation achieved 0.1\% error, compared to the ground truth.  


The results in Table~\ref{table:RQ1} show that Exponentiality testing as a stopping criterion meets our conditions for accurate predictions in 17 cases whereas Laplace led to an accurate prediction of the ground truth in 13 cases (both out of 20 cases). Within the Exponentiality testing, we study the performance of bootstrapping vs. Quantile methods to pick threshold of GEV accurately. We observe that bootstrapping achieves better results than the Quantile method in 3 cases. 
Finally, we compare two types of GEV distributions, i.e., exponential vs. PP distributions. Our results show that both techniques have similar performance w.r.t. the best prediction. 
While the lowest error across all the different configurations that used Exponentiality testing, bootstrapping, and exponential distribution of GEV is +0.26\%; the lowest error of benchmarks that used Exponentiality testing, bootstrapping, and Poisson Process distribution is +0.1\%. 



\begin{tcolorbox}[boxrule=0.5pt,left=0.5pt,right=0.5pt,top=0.5pt,bottom=0.5pt]
\textbf{Answer RQ1:}
First, we find that Exponentiality testing, as a stopping criterion during differential fuzzing, leads to more accurate EVT-based extrapolations. Second, our experiments show that bootstrapping (as a method for inferring the threshold of extreme values) with Poisson Process (as a type of GEV distribution) slightly outperforms Quantile and exponential distribution.  
\end{tcolorbox}

\input{Sections/Table_RQ2}
\subsection{RQ2 -- Comparing the EVT to the baseline}

The previous experiments convinced us that the Exponentiality stopping criterion, bootstrapping, and PP distribution is one of the best configurations to predict the worst-case cost differences via EVT. In this section, we compare our extrapolation approach via EVT to three baseline statistical techniques; Markov's Inequality~\cite{chernoff1952measure}, Chebyshev’s Inequality~\cite{chebyshev1867valeurs}, and Bayes Factor (following Jeffery's Test~\cite{jha2009bayesian}).
Following the standard requirements to achieve a confidence of $1 - c$ with an error probability below $\alpha$ of estimation for concentration inequalities, we need a training sample size of $O(\frac{ln(1/c)}{\alpha})$. We set the training size to at least $1,200$ to achieve a 0.95 confidence with an error probability 0.05. 
Following common practices, we used Laplace testing to determine the sample size for Bayes Factor. 

Table~\ref{table-RQ2-results} show the results of comparing the EVT extrapolations to Markov, Chebyshev, and Bayes factor. The statistically significant results based on Mann-Whitney U-test are highlighted in bold. When there is more than one highlighted row for a benchmark, it means two or more techniques are tied. Overall, EVT-based extrapolations are highlighted in 11 cases out of 14 benchmarks while Bayes factor is highlighted in 10 cases out of 14 benchmarks. The Chebyshev is highlighted in 7 cases out of 14 benchmarks, but Markov did not win in any cases. When comparing EVT to Bayes factor, we observe that EVT wins in 2 cases, losses in 1, and ties in 9 cases. However, the winning case for the Bayes factor (i.e., Leak Set 1) is an underestimation that may still miss bugs. Comparing EVT to Chebyshev, we see that EVT wins, losses, and ties in 5, 2, and 5 cases. 

Specifically, both EVT-based extrapolations and Bayes factor perform very well for 3 cases of ``Array Unsafe", ``Sanity Unsafe", and ``Straightline Unsafe" (0\% error); but Chebyshev has some errors in all benchmarks. 
The dynamic training sizes used by the Bayes factor (via Laplace testing) and EVT (via Exponentiality testing) seem more effective than the fixed training size for Markov/Chebyshev.
Since tight over-approximations are preferred, we observe that in 5 out of 11 remaining benchmarks, EVT-based extrapolations provide a tighter over-approximation than Bayes factor. There are no cases where Bayes factor outperforms EVT-based extrapolations in terms of tight over-approximation. 


\begin{tcolorbox}[boxrule=0.5pt,left=0.5pt,right=0.5pt,top=0.5pt,bottom=0.5pt]
\textbf{Answer RQ2:}
EVT-based extrapolations outperform the most competitive baseline (Bayes Factor via Jeffery's Test~\cite{jha2009bayesian}). In 57.1\% of the cases, EVT-based extrapolations provide a tight over-approximation, while the Bayes Factor under-approximates the ground truth in 78.6\% of cases. 
\end{tcolorbox}

\input{Sections/Table_RQ3}

\subsection{RQ3 -- Measuring the accuracy and performance gain of EVT-enabled Differential Fuzzing in realistic Java libraries}
Table~\ref{tab-RQ3} shows the results of EVT-enabled differential fuzzing in larger Java benchmarks.
The positive errors in all cases show that EVT-based extrapolations provide over-approximation, so it does not underestimate the worst-case cost differences. The error ranges from 32.8\% (Stateless Authenticated) to 206.8\% (Apache FtpServer Stringutils). In cases when maximum of unobserved test data is higher than the maximum of observed training data, the EVT prediction provides very close prediction (e.g., 51 vs. 52 for Jetty, 341 vs 381 for Tourplanner, and 143 vs 175 for Apache Ftpserver).

When the max. cost difference is the same between observed training and unseen testing data, the EVT prediction still results in an over-approximation (positive error). This means that even when the maximum observed cost during training is the true maximum, the EVT model is still projecting a tail that extends beyond this observed maximum. This is inherent to how EVT models tail behavior; it extrapolates beyond the observed data.

Finally, RQ3 includes larger-scale Java libraries such as \textit{Spring Security} and \textit{Apache Ftpserver}. 
We also calculate the performance gain of EVT-enabled differential fuzzing over baseline methods in terms of bytecode execution saved by early stopping of fuzzing campaigns. We report the performance gain of early return in the last column of Table~\ref{tab-RQ3}. The results shows that in one case for \textit{Apache Ftpserver Salted}, 1,674,774,946 bytecode executions has been saved. Since we run all the benchmarks for 30 mins, the ratio of $\frac{|Testing|}{|Testing|+|Training|}$ approximately shows the wall clock savings (e.g., 0.7*30 $\approx$ 21 mins saved for Stateless Auth). 
Finding a trade-off between the accuracy of the prediction (the "Error" column) vs. the performance gain is an interesting direction for future work. 

Finally, for those cases with larger errors in the prediction, we also notice that the Scale parameter of the EVT distribution is large. This can be used to guide a search algorithm to potentially find better threshold of extreme values.

\begin{tcolorbox}[boxrule=0.5pt,left=0.5pt,right=0.5pt,top=0.5pt,bottom=0.5pt]
\textbf{Answer RQ3:}
We find that EVT-based differential fuzzing does not underestimate the worst-case cost differences in any larger Java libraries. The error ranges from 32.8\% (Stateless Authenticated) to 206.8\% (Apache FtpServer Stringutils). In 4 out of 9 cases, EVT-based differential fuzzing provides a tight upper-bound of the worst-case cost differences. 
We also report the significant performance gain of fuzzing when early stopping and extrapolation via EVT is applied. 
\end{tcolorbox}

%% file: Sections/Table_RQ1.tex
\begin{table*}
\centering
\caption{Comparing statistical methods for early stopping of fuzzing (Exponentiality vs. Laplace), selecting the threshold of GEV distributions (Bootstrap vs. Quantile), and the type of GEV distributions (Exponential vs. Poisson Process Distributions). 
}
\resizebox{\textwidth}{!}{%
\begin{tabu}{|c c | c c c c c c c c | c c c c c c c c |}
\hline

 &  & \multicolumn{8}{c|}{\textbf{GEV (Exponentiality Testing)}} & \multicolumn{8}{c|}{\textbf{GEV (Laplace Testing)}}
\\ 
Benchmark & Num. Inputs & Threshold & Type & 
Training & Testing & Ground Truth & Prediction & Best(Prediction) & Best(Error\%) 
& Threshold & Type & 
Training & Testing & Ground Truth & Prediction & Best(Prediction) & Best(Error\%)
\\
\hline
\multirow{4}{2.5em}{Leak Set (1)} & \multirow{4}{3.5em}{13,186 (+/- 369)} & \multirow{2}{2.5em}{Bootstrap} & 
\textbf{Exp.} & \textbf{7.7k (+/-5.2k)} & \textbf{5.8k (+/-4.2k)} & \textbf{1.1k (+/-21.1)} & \textbf{1.8k (+/-0.7k)} & \textbf{1.1k} & \textbf{2.54} & 
\multirow{2}{2.5em}{Bootstrap} & 
Exp. & 107.3 (+/-23.0) & 14.3k (+/-0.4k) & 1.1k (+/-23.14) & 3.4k (+/-2.1k) & 629.0 & 41.1\\ & &&
\textbf{PP} & \textbf{7.7k (+/-5.3k)} & \textbf{5.9k (+/-4.2k)} & \textbf{1.1k (+/-21.4)} & \textbf{1.1k (+/-0.3k)} & \textbf{1.1k} & \textbf{0.1} && 
PP & 110 (+/- 26.5) & 14.9k (+/-4.7k) & 1.1k (+/-25.5) & 1.3k (+/-1.2k) & 841.6 & 21.2 \\ & &
\multirow{2}{2.5em}{Quantile} & 
\textbf{Exp.} & \textbf{2.0k (+/-0.6k)} & \textbf{1.2k (+/-2.7k)} & \textbf{1.1k (+/-21.1)} & \textbf{2.3k (+/-1.1k)} & \textbf{1.1k} & \textbf{3.69} & 
\multirow{2}{2.5em}{Quantile} & 
Exp. & 0.1k (+/-15.2) & 13.5k (+/-3.0k) & 1.1k (+/-17.6) & 3.9k (+/-2.2k) & 1.3k & 20.0\\ & &&
\textbf{PP} & \textbf{2.0k (+/-0.6k)} & \textbf{11.5k (+/-2.8k)} & \textbf{1.1 (+/-20.9)} & \textbf{1.0k (+/-0.3k)} & \textbf{1.1k} & \textbf{1.86} && 
\textbf{PP} & \textbf{100 (+/-0)} & \textbf{12.9k (+/-0.4k)} & \textbf{1.1k (+/-19.0)} & \textbf{1.0k (+/-0.5k)}
 & \textbf{1.1k} & \textbf{3.1} \\

\hline

\multirow{4}{2.5em}{Leak Set (2)} & \multirow{4}{3.5em}{13,371 (+/- 168)} & 
\multirow{2}{2.5em}{Bootstrap} & 
\textbf{Exp.} & \textbf{6.6k (+/-5.9k)} & \textbf{7.0k (+/-5.0k)} & \textbf{1.4k (+/-19.91)} & \textbf{2.1k (+/-0.8)} & \textbf{1.5k} & \textbf{1.72} &\multirow{2}{2.5em}{Bootstrap} & 
\textbf{Exp.} & \textbf{100(+/-0)} & \textbf{12.9k (+/-0.3k)} & \textbf{1.5k (+/-13.8)} & \textbf{2.6k (+/-1.3k)} & \textbf{1.5k} & \textbf{4.92} \\ & &&
\textbf{PP} & \textbf{6.6k (+/-5.9k)} & \textbf{7.0k (+/-5.0k)} & \textbf{1.5k (+/-19.91)} & \textbf{1.4k (+/-0.4k)} & \textbf{1.5k} & \textbf{0.32} &&
\textbf{PP} & \textbf{100 (+/-0)} & \textbf{12.9k (+/-0.3k)} & \textbf{1.5k (+/-13.8)} & \textbf{9.9k (+/-25k)} & \textbf{1.5k} & \textbf{2.65} \\ & &
\multirow{2}{2.5em}{Quantile} & 
Exp. & 1.4k (+/-0.3k) & 12.2k (+/-2.4k) & 1.5k (+/-19.91) & 3.7k (+/-0.9k) & 1.3k & 13.02 &\multirow{2}{2.5em}{Quantile} & 
\textbf{Exp.} & \textbf{108.5 (+/-33.4)} & \textbf{13.5k (+/-2.7k)} & \textbf{1.5k (+/- 22.9)} & \textbf{4.4k (+/-1.6k)} & \textbf{1.5k} & \textbf{1.58} \\ & &&
\textbf{PP} & \textbf{1.4k (+/-0.3k)} & \textbf{12.2k (+/-2.5k)} & \textbf{1.5k (+/-19.91)} & \textbf{1.1k (+/-0.4k)} & \textbf{1.5k} & \textbf{8.37} &&
PP & 112 (+/- 39.1) & 13.7k (+/- 3.2k) & 1.5k (+/-25.4) & 42.k (+/-155.2k) & 1157 & 18.75\\

\hline

\multirow{4}{2.5em}{Leak Set (3)} & \multirow{4}{3.5em}{13,434 (+/- 307)} & 
\multirow{2}{2.5em}{Bootstrap} & 
\textbf{Exp.} & \textbf{9.7k (+/-13.5k)} & \textbf{5.8k (+/-5.0k)} & \textbf{1.7k (+/-82.6)} & \textbf{2.1k (+/-0.8k)} & \textbf{1.6k} & \textbf{1.06} & \multirow{2}{2.5em}{Bootstrap} & 
\textbf{Exp.} & \textbf{104.2 (+/-13.4)} & \textbf{13.2k (+/-0.3k)} & \textbf{1.7k (+/-80.9)} & \textbf{4.5k (+/-2.0k)} & \textbf{1.9k} & \textbf{9.03}\\ & &&
\textbf{PP} & \textbf{10.0k (+/-13.6k)} & \textbf{5.6k (+/-4.9k)} & \textbf{1.7k (+/-83.9)} & \textbf{1.5k (+/-0.3k)} & \textbf{1.6k} & \textbf{2.99}&&
\textbf{PP} & \textbf{105.6 (+/-15.3)} & \textbf{13.2k (+/-0.3k)} & \textbf{1.7k (+/-84.3)} & \textbf{462.2k (+/- 1.0m)} & \textbf{1.7k} & \textbf{2.47}\\ & &
\multirow{2}{2.5em}{Quantile} & 
Exp. & 1.3k (+/-0.3k) & 14.3k (+/-11.7k) & 1.7k (+/-84.2) & 3.1k (+/-847.86) & 1.6k & 19.71&\multirow{2}{2.5em}{Quantile} & 
{Exp.} & {103.6 (+/- 11.7)} & {13.2k (+/-0.2k)} & {1.7k (+/-84.8)} & {4.4k (+/-2.0k)} & {1.5k} & 11.4\\ & &&
\textbf{PP} & \textbf{1.3k (+/- 0.3k)} & \textbf{14.3k (+/-11.9k)} & \textbf{1.7k (+/-83.9)} & \textbf{1.3k (+/-0.3k)} & \textbf{1.8k} & \textbf{3.4}&&
\textbf{PP} & \textbf{104.1 (+/- 12.3)} & \textbf{13.2k (+/-0.2k)} & \textbf{1.7k (+/-85.5)} & \textbf{21.0k (+/-0.7k)} & \textbf{1.7k} & \textbf{2.15}\\

\hline

\multirow{4}{2.5em}{Leak Set (4)} & \multirow{4}{3.5em}{13,157 (+/- 442)} & 
\multirow{2}{2.5em}{Bootstrap} & 
\textbf{Exp.} & \textbf{6.9k (+/-5.2k)} & \textbf{6.1k (+/-5.2k)} & \textbf{2.1k (+/-0.1k)} & \textbf{2.3k (+/-0.8k)} & \textbf{2.2k} & \textbf{1.38} &\multirow{2}{2.5em}{Bootstrap} & 
\textbf{Exp.} & \textbf{0.1k (+/- 26.8)} & \textbf{13.0k (+/-0.4k)} & \textbf{2.1k (+/-87.8)} & \textbf{4.6k (+/-1.8k)} & \textbf{2.2k} & \textbf{5.56}\\ & &&
\textbf{PP} & \textbf{6.8k (+/-5.3k)} & \textbf{6.3k (+/-5.2k)} & \textbf{2.1k (+/-0.1k)} & \textbf{2.0k (+/-0.9k)} & \textbf{2.2k} & \textbf{1.06}&&
PP & 0.1k (+/-30.5) & 13.0k (+/-0.4k) & 2.1k (+/-63.0) & 9.3k (+/-12.5k) & 3.0k & 37.07\\ & &
\multirow{2}{2.5em}{Quantile} & 
\textbf{Exp.} & \textbf{1.1k (+/-0.2k)} & \textbf{12.0k (+/-0.4k)} & \textbf{2.1k (+/-0.1k)} & \textbf{3.0k (+/-0.8k)} & \textbf{2.1k} & \textbf{0.53} &\multirow{2}{2.5em}{Quantile} & 
\textbf{Exp.} & \textbf{0.1k (+/- 0)} & \textbf{13.1k (+/- 0.4k)} & \textbf{2.1k (+/- 0.1k)} & \textbf{6.1k (+/-1.7k)} & \textbf{2.2k} & \textbf{1.78}\\ & &&
PP & 1.1k (+/-0.2k) & 12.0k (+/-0.4k) & 2.1k (+/-0.1k) & 1.7k (+/-0.7k) & 1.9k & 13.9 &&
\textbf{PP} & \textbf{100 (+/- 0)} & \textbf{13.1k (+/-0.4k)} & \textbf{2.1k (+/-0.1k)} & \textbf{6.1k (+/-1.7k)} & \textbf{2.2k} & \textbf{1.78}\\

\hline

\multirow{4}{2.5em}{Leak Set (5)} & \multirow{4}{3.5em}{13,205 (+/- 456)} & 
\multirow{2}{2.5em}{Bootstrap} & 

\textbf{Exp.} & \textbf{6.4k (+/-5.3k)} & \textbf{6.6k (+/-5.5k)} & \textbf{2.4k (+/-0.1k)} & \textbf{2.7k(+/-0.7k)} & \textbf{2.4k} & \textbf{0.26} & \multirow{2}{2.5em}{Bootstrap} & 
\textbf{Exp.} & \textbf{0.1k (+/- 13.6)} & \textbf{12.9k (+/- 0.4k)} & \textbf{2.4k (+/- 0.1k)} & \textbf{5.3k (+/- 3.4k)} & \textbf{2.4k} & \textbf{3.89}\\ & &&
\textbf{PP} & \textbf{6.4k (+/-5.3k)} & \textbf{6.6k (+/-5.5k)} & \textbf{2.4k (+/-0.1k)} & \textbf{2.7k (+/-0.7k)} & \textbf{2.4k} & \textbf{2.64}&&

PP & 0.1k (+/-16.5) & 12.9k (+/- 0.5k) & 2.4k (+/- 0.1k) & 5.3k (+/- 3.4k) & 2.4k & 24.93\\ & &
\multirow{2}{2.5em}{Quantile} & 
\textbf{Exp.} & \textbf{0.9k (+/-0.2k)} & \textbf{12.1k (+/-0.4k)} & \textbf{2.4k (+/-0.1k)} & \textbf{2.8k (+/-0.4k)} & \textbf{2.6k} & \textbf{4.74} &\multirow{2}{2.5em}{Quantile} & 
\textbf{Exp.} & \textbf{0.1k (+/- 0)} & \textbf{12.9k (+/- 0.4k)} & \textbf{2.4k (+/- 0.1k)} & \textbf{5.5k (+/- 2.5k)} & \textbf{2.3k} & \textbf{5.08}\\ & &&
\textbf{PP} & \textbf{0.9k (+/-0.2k)} & \textbf{12.1k (+/- 0.4k)} & \textbf{2.4k (+/-0.1k)} & \textbf{1.9k (+/-0.7k)} & \textbf{2.4k} & \textbf{0.42}&&
\textbf{PP} & \textbf{100 (+/- 0)} & \textbf{12.9k (+/- 0.4k)} & \textbf{2.4k (+/-0.1k)} & \textbf{8.4k (+/-20.4k)} & \textbf{2.2k} & \textbf{7.62}\\

\hline

\end{tabu}
}
\label{table:RQ1}
\end{table*}

%% file: Sections/Table_RQ2.tex
\begin{table*}
\centering
 \caption{Comparison of EVT-based extrapolations with the baseline methods in predicting the worst-case cost differences.
 The highlighted values are winners based on the Mann–Whitney U-test.}
\resizebox{0.99\textwidth}{!}{
 \begin{tabular}{ | c  c  | c  c  c  c  c  c  c | } 
  \hline
   
 \textbf{Benchmark} & Num. Inputs & Extrapolation & $|$Training$|$ & Max. Training & $|$Testing$|$  & Max. Testing & Prediction & Error \% \\ 
 \hline
 

\multirow{2}{6.0em}{Leak Set (1)} & \multirow{2}{9.0em}{13186.0 (+/- 369.2)} & 
Markov & 1,200 (+/- 0.0) & 744.2 (+/- 33.41) & 11985.0 (+/- 369.2) & 1089.6 (+/- 19.7) & 7266.0 (+/- 4561.2) & 565.2 (+/- 410.32)  \\
& & {Chebyshev} & {1,200 (+/- 0.0)} & {744.2 (+/- 33.4)} & {11985.0 (+/- 369.2)} & {1089.6 (+/- 19.7)} & {1660.9 (+/- 171.3)} & {52.5 (+/- 16.0)}  \\  
& & \textbf{Bayes Factor} & \textbf{6114.0 (+/- 1864.4)} & \textbf{1039.0 (+/- 41.4)} & \textbf{19774.2 (+/- 5365.2)} & \textbf{1089.6 (+/- 19.7)} & \textbf{1039.0 (+/- 41.4)} & \textbf{-4.58 (+/- 5.08)}  \\ 
& & {EVT} & {8999.2 (+/- 3155.4)} & {1075.2 (+/- 16.1)} & {4211.2 (+/- 3159.0)} & {1089.6 (+/- 19.7)} & {1240.5 (+/- 507.3)} & {13.5 (+/- 45.2)}  \\
\hline 

\multirow{2}{6.0em}{Leak Set (2)} & \multirow{2}{9.0em}{13371.2 (+/- 167.7)} & 
Markov & 1,200 (+/- 0.0) & 975.2 (+/- 123.4) & 12170.2 (+/- 167.7) & 1462.4 (+/- 21.5) & 12673.7 (+/- 1461.7) & 766.66 (+/- 98.73)  \\
& & \textbf{Chebyshev} & \textbf{1,200 (+/- 0.0)} & \textbf{975.2 (+/- 123.4)} & \textbf{12170.2 (+/- 167.7)} & \textbf{1462.4 (+/- 21.5)} & \textbf{1457.6 (+/- 280.9)} & \textbf{-0.31 (+/- 19.15)}  \\
& & \textbf{Bayes Factor} & \textbf{5706.0 (+/- 980.7)} & \textbf{1299.4 (+/- 148.9)} &  \textbf{18910.0 (+/- 3348.2)} & \textbf{1462.4 (+/- 21.5)} & \textbf{1299.4 (+/- 148.9)} & \textbf{-11.11 (+/- 10.39)}  \\ 
& & \textbf{EVT} & \textbf{8201 (+/- 1795.4)} & \textbf{1414.8 (+/- 68.6)} & \textbf{5168.5 (+/- 1963.4)} & \textbf{1460 (+/- 24)} & \textbf{1505.3 (+/- 121.2)} & \textbf{3.2 (+/- 9.2)} \\

\hline 

\multirow{2}{6.0em}{Leak Set (3)} & \multirow{2}{9.0em}{13433.6 (+/- 307.3)} & 
Markov & 1,200 (+/- 0.0) & 1214.4 (+/- 77.0) & 12232.6 (+/- 307.3) & 1717.6 (+/- 108.6) & 37161.3 (+/- 1477.3) & 2068.79 (+/- 129.7)  \\
& & \textbf{Chebyshev} & \textbf{1,200 (+/- 0.0)} & \textbf{1214.4 (+/- 77.0)} & \textbf{12232.6 (+/- 307.3)} & \textbf{1717.6 (+/- 108.6)} & \textbf{1606.9 (+/- 342.6)} & \textbf{-5.99 (+/- 22.24)}  \\  
& & \textbf{Bayes Factor} & \textbf{7440.0 (+/- 1834.7)} & \textbf{1425.0 (+/- 129.2)} & \textbf{24387.4 (+/- 6529.6)} & \textbf{1717.6 (+/- 108.6)} & \textbf{1425.0 (+/- 129.2)} & \textbf{-16.86 (+/- 7.9)}  \\ 
& & \textbf{EVT} & \textbf{5766 (+/- 2522.6)} & \textbf{1525.8 (+/- 151.2)} & \textbf{7691.4 (+/- 2735.2)} & \textbf{1717.6 (+/- 108.6)} & \textbf{1508.4 (+/- 218.1)} & \textbf{-12.45 (+/- 8.9)}  \\ 
\hline 

\multirow{2}{6.0em}{Leak Set (4)} & \multirow{2}{9.0em}{13156.8 (+/- 442.1)} & 
Markov & 1,200 (+/- 0.0) & 1380.0 (+/- 243.4) & 11955.8 (+/- 442.1) & 2189.6 (+/- 41.1) & 55868.0 (+/- 1719.0) & 2452.42 (+/- 97.72)  \\
& & \textbf{Chebyshev} & \textbf{1,200 (+/- 0.0)} & \textbf{1380.0 (+/- 243.4)} & \textbf{11955.8 (+/- 442.1)} & \textbf{2189.6 (+/- 41.1)} & \textbf{1964.6 (+/- 608.4)} & \textbf{-10.31 (+/- 27.38)}  \\  
& & \textbf{Bayes Factor} & \textbf{5808.0 (+/- 524.2)} & \textbf{1720.0 (+/- 268.0)} & \textbf{18720.0 (+/- 2103.9)} & \textbf{2189.6 (+/- 41.1)} & \textbf{1720.0 (+/- 268.0)} & \textbf{-21.54 (+/- 11.41)}  \\ 
& & \textbf{EVT} & \textbf{8791.4 (+/- 2171.1)} & \textbf{2047.0 (+/- 39.8)} & \textbf{4392.6 (+/- 1822.9)} & \textbf{2189.6 (+/- 41.1)} & \textbf{2129.4 (+/- 559.6)} & \textbf{-2.83 (+/- 24.96)}  \\ 
\hline 

\multirow{2}{6.0em}{Leak Set (5)} & \multirow{2}{9.0em}{13205.0 (+/- 456.1)} & 
Markov & 1,200 (+/- 0.0) & 1711.2 (+/- 167.1) & 12004.0 (+/- 456.1) & 2412.0 (+/- 121.2) & 79846.01 (+/- 1538.7) & 3218.5 (+/- 210.06)  \\
& & \textbf{Chebyshev} & \textbf{1,200 (+/- 0.0)} & \textbf{1711.2 (+/- 167.1)} & \textbf{12004.0 (+/- 456.1)} & \textbf{2412.0 (+/- 121.2)} & \textbf{2527.8 (+/- 280.5)} & \textbf{4.75 (+/- 9.47)}  \\  
& & Bayes Factor & 6252.0 (+/- 1015.5) & 1748.0 (+/- 130.1) & 20149.8 (+/- 2527.6) & 2412.0 (+/- 121.2) & 1748.0 (+/- 130.1) & -27.22 (+/- 8.73)  \\ 
& & {EVT} & {9741.2 (+/- 2058.2)} & {2208 (+/- 195.2)} & {3496.4 (+/- 2350.5)} & {2412 (+/- 121.2)} & {2014.8 (+/- 172.0)} & {-16.48 (+/- 5.43)}  \\
\hline 

\multirow{2}{6.0em}{Array Unsafe} & \multirow{2}{9.0em}{10567.8 (+/- 313.8)} & 
Markov & 1,200 (+/- 0.0) & 192.0 (+/- 0.0) & 9366.8 (+/- 313.8) & 192.0 (+/- 0.0) & 4489.1 (+/- 821.2) & 2238.05 (+/- 427.71)  \\
& & Chebyshev & 1,200 (+/- 0.0) & 192.0 (+/- 0.0) & 9366.8 (+/- 313.8) & 192.0 (+/- 0.0) & 853.4 (+/- 56.7) & 344.47 (+/- 29.52)  \\  
& & \textbf{Bayes Factor} & \textbf{4410.0 (+/- 458.4)} & \textbf{192.0 (+/- 0.0)} & \textbf{10833.4 (+/- 1387.3)} & \textbf{192.0 (+/- 0.0)} & \textbf{192.0 (+/- 0.0)} & \textbf{0.0 (+/- 0.0)}  \\ 
& & \textbf{EVT} & \textbf{6955 (+/- 1004.1)} & \textbf{192.0 (+/- 0.0)} & \textbf{3797 (+/- 783.5)} & \textbf{192 (+/- 0.0)} & \textbf{192 (+/- 0.0)} & \textbf{0.0 (+/- 0.0)}  \\ 

\hline 

\multirow{2}{6.0em}{gpt14 Unsafe} & \multirow{2}{9.0em}{13156.0 (+/- 226.9)} & 
Markov & 1,200 (+/- 0.0) & 4692.4 (+/- 1229.2) & 11955.0 (+/- 226.9) & 6389.6 (+/- 455.6) & 181668.2 (+/- 2512.1) & 2755.44 (+/- 219.03)  \\
& & \textbf{Chebyshev} & \textbf{1,200 (+/- 0.0)} & \textbf{4692.4 (+/- 1229.2)} & \textbf{11955.0 (+/- 226.9)} & \textbf{6389.6 (+/- 455.6)} & \textbf{7284.5 (+/- 783.0)} & \textbf{14.21 (+/- 11.53)}  \\  
& & \textbf{Bayes Factor} & \textbf{5898.0 (+/- 920.4)} & \textbf{5470.2 (+/- 548.0)} & \textbf{18927.8 (+/- 2430.7)} & \textbf{6389.6 (+/- 455.6)} & \textbf{5470.2 (+/- 548.0)} & \textbf{-14.42 (+/- 5.28)}  \\
& & \textbf{EVT} & \textbf{9266.6 (+/- 3785.6)} & \textbf{5795.6 (+/- 469.8)} & \textbf{4210.8 (+/- 3855.4)} & \textbf{6389.6 (+/- 455.6)} & \textbf{5257.2 (+/- 682.3)} & \textbf{-17.12 (+/- 14)}  \\ 
\hline 

\multirow{2}{6.0em}{k96 Unsafe} & \multirow{2}{9.0em}{13596.8 (+/- 954.8)} & 
Markov & 1,200 (+/- 0.0) & 3867.6 (+/- 743.7) & 12395.8 (+/- 954.8) & 5291.8 (+/- 577.1) & 83100.2 (+/- 2444.8) & 1485.42 (+/- 180.95)  \\
& & \textbf{Chebyshev} & \textbf{1,200 (+/- 0.0)} & \textbf{3867.6 (+/- 743.7)} & \textbf{12395.8 (+/- 954.8)} & \textbf{5291.8 (+/- 577.1)} & \textbf{3680.6 (+/- 459.2)} & \textbf{-29.92 (+/- 10.42)}  \\  
& & \textbf{Bayes Factor} & \textbf{5220.0 (+/- 1014.5)} & \textbf{3939.6 (+/- 717.9)} & \textbf{17812.0 (+/- 3640.0)} & \textbf{5291.8 (+/- 577.1)} & \textbf{3939.6 (+/- 717.9)} & \textbf{-25.85 (+/- 9.19)}  \\ 
& & \textbf{EVT} & \textbf{10177.4 (+/- 1455.9)} & \textbf{4321.2 (+/- 275.9)} & \textbf{3667.4 (+/- 1732.1)} & \textbf{5291.8 (+/- 577.1)} & \textbf{6723.9 (+/- 6050.7)} & \textbf{25.29 (+/- 105.26)}  \\ 
\hline 

\multirow{2}{6.0em}{login Unsafe} & \multirow{2}{9.0em}{12804.6 (+/- 389.4)} & 
Markov & 1,200 (+/- 0.0) & 6.4 (+/- 6.1) & 11603.6 (+/- 389.4) & 62.0 (+/- 0.0) & 93.9 (+/- 86.8) & 51.38 (+/- 140.07)  \\
& & Chebyshev & 1,200 (+/- 0.0) & 6.4 (+/- 6.1) & 11603.6 (+/- 389.4) & 62.0 (+/- 0.0) & 17.0 (+/- 13.6) & -72.56 (+/- 21.9)  \\  
& & Bayes Factor & 6660.0 (+/- 901.0) & 12.0 (+/- 8.0) & 20684.8 (+/- 3214.7) & 62.0 (+/- 0.0) &  12.0 (+/- 8.0) & -80.65 (+/- 12.9))  \\ 
& & \textbf{EVT} & \textbf{6670.8 (+/- 987.3)} & \textbf{61.2 (+/- 1.8)} & \textbf{6135.8 (+/- 1047.4)} & \textbf{62 (+/- 0)} & \textbf{84.1 (+/- 18.8)} & \textbf{35.64 (+/- 30.31)} \\ 
\hline 

\multirow{2}{6.0em}{modPow1 Unsafe} & \multirow{2}{9.0em}{12392.8 (+/- 352.2)} & 
Markov & 1,200 (+/- 0.0) & 1624.0 (+/- 246.1) & 11191.8 (+/- 352.2) & 2543.2 (+/- 328.6) & 38153.3 (+/- 1728.0) & 1412.79 (+/- 122.83)  \\
& & \textbf{Chebyshev} & \textbf{1,200 (+/- 0.0)} & \textbf{1624.0 (+/- 246.07)} & \textbf{11191.8 (+/- 352.2)} & \textbf{2543.2 (+/- 328.6)} & \textbf{2359.0 (+/- 320.9)} & \textbf{-6.66 (+/- 12.68)}  \\  
& & Bayes Factor & 5946.0 (+/- 795.5) & 2002.4 (+/- 157.0) & 17510.0 (+/- 1902.8) & 2543.2 (+/- 328.6) &  2002.4 (+/- 157.0) & -20.03 (+/- 13.53)  \\ 
& & EVT & 8234.8 (+/- 1489.4) & 2214.8 (+/- 107.0) & 4235.4 (+/- 1149.7) & 2543.2 (+/- 328.6) & 2034.6 (+/- 162.1) & -19.37 (+/- 8.18)  \\ 
\hline

\multirow{2}{6.0em}{modPow2 Unsafe} & \multirow{2}{9.0em}{14782.0 (+/- 2535.4)} & 
Markov & 1,200 (+/- 0.0) & 36.8 (+/- 46.4) & 13581.0 (+/- 2535.4) & 143.8 (+/- 11.3) & 46.5 (+/- 63.1) & -68.24 (+/- 42.65)  \\
& & Chebyshev & 1,200 (+/- 0.0) & 36.8 (+/- 46.4) & 13581.0 (+/- 2535.4) & 143.8 (+/- 11.3) & 38.3 (+/- 49.8) & -73.79 (+/- 33.63)  \\  
& & Bayes Factor & 7014.0 (+/- 2442.9) & 81.2 (+/- 49.3) & 27784.8 (+/- 16389.9) & 143.8 (+/- 11.3) &  81.2 (+/- 49.3) & -44.70 (+/- 31.05)  \\ 
& & \textbf{EVT} & \textbf{6652.6 (+/- 3457.4)} & \textbf{121.6 (+/- 17.6)} & \textbf{8143 (+/- 5511.8)} & \textbf{143.8 (+/- 11.3)} & \textbf{135.2 (+/- 57.2)} & \textbf{-6.87 (+/- 36.41)}  \\ 
\hline

\multirow{2}{6.0em}{passwordEq Unsafe} & \multirow{2}{9.0em}{12719.6 (+/- 172.0)} & 
Markov & 1,200 (+/- 0.0) & 56.0 (+/- 15.6) & 11518.6 (+/- 172.0) & 63.0 (+/- 0.0) & 2175.5 (+/- 670.6) & 3353.18 (+/- 1064.47)  \\
& & Chebyshev & 1,200 (+/- 0.0) & 56.0 (+/- 15.6) & 11518.6 (+/- 172.01) & 63.0 (+/- 0.0) & 239.1 (+/- 82.6) & 279.56 (+/- 131.13)  \\  
& & \textbf{Bayes Factor} & \textbf{5532.0 (+/- 1287.8)} & \textbf{60.8 (+/- 4.9)} & \textbf{17368.0 (+/- 4021.1)} & \textbf{63.0 (+/- 0.0)} &  \textbf{60.8 (+/- 4.9)} & \textbf{-3.5 (+/- 7.81)}  \\ 
& & \textbf{EVT} & \textbf{5798 (+/- 258.8)} & \textbf{63 (+/- 0)} & \textbf{6816 (+/- 188.1)} & \textbf{63 (+/- 0)} & \textbf{69.6 (+/- 9.4)} & \textbf{10.52 (+/- 14.9)}  \\ 
\hline

\multirow{2}{6.0em}{sanity Unsafe} & \multirow{2}{9.0em}{1464.4 (+/- 234.1)} & 
Markov & 1,200 (+/- 0.0) & 2867086184.5 (+/- 351155148.48) & 329.8 (+/- 209.1) & 2774479762.0 (+/- 419417121.7) & 80645893388.6 (+/- 17819138629.2) & 2606.11 (+/- 971.09)  \\
& & Chebyshev & 1,200 (+/- 0.0) & 2867086184.5 (+/- 351155148.5) & 329.8 (+/- 209.1) & 2774479762.0 (+/- 419417121.7) & 12611863196.6 (+/- 832399251.3) & 317.05 (+/- 81.37)  \\  
& & \textbf{Bayes Factor} & \textbf{2234.0 (+/- 333.6)} & \textbf{2806659155.4 (+/- 332775574.5)} & \textbf{0.0 (+/- 0.0)} & \textbf{2806659155.4 (+/- 332775574.5)} &  \textbf{2806659155.4 (+/- 332775574.5)} & \textbf{0.0 (+/- 0.0)}  \\ 
& & \textbf{EVT} & \textbf{2234.0 (+/- 333.6)} & \textbf{2806659155.4 (+/- 332775574.5)} & \textbf{0.0 (+/- 0.0)} & \textbf{2806659155.4 (+/- 332775574.5)} & \textbf{2806659155.4 (+/- 332775574.5)} & \textbf{0.0 (+/- 0.0)}  \\ 
\hline

\multirow{2}{6.0em}{straightline Unsafe} & \multirow{2}{9.0em}{14207.2 (+/- 96.0)} & 
Markov & 1,200 (+/- 0.0) & 8.0 (+/- 0.0) & 13006.2 (+/- 96.0) & 8.0 (+/- 0.0) & 152.8 (+/- 65.3) & 1810.07 (+/- 816.47)  \\
& & Chebyshev & 1,200 (+/- 0.0) & 8.0 (+/- 0.0) & 13006.2 (+/- 96.0) & 8.0 (+/- 0.0) & 31.7 (+/- 8.2) & 296.09 (+/- 103.02)  \\  
& & \textbf{Bayes Factor} & \textbf{4788.0 (+/- 989.0)} & \textbf{8.0 (+/- 0.0)} & \textbf{17432.8 (+/- 3461.0)} & \textbf{8.0 (+/- 0.0)} &  \textbf{8.0 (+/- 0.0)} & \textbf{0.0 (+/- 0.0)}  \\ 
& & \textbf{EVT} & \textbf{4788.0 (+/- 989.0)} & \textbf{8.0 (+/- 0.0)} & \textbf{17432.8 (+/- 3461.0)} & \textbf{8.0 (+/- 0.0)} & \textbf{8.0 (+/- 0.0)} & \textbf{0.0 (+/- 0.0)}  \\ 
\hline

 \end{tabular}
 }
 \label{table-RQ2-results}
\end{table*}

%% file: Sections/Table_RQ3.tex
\begin{table*}
\centering
 \caption{Evaluation of EVT-enabled \dfuzz in larger Java libraries.}
\resizebox{0.95\textwidth}{!}{
 \begin{tabular}{ | c  c  | c  c  c  c  c c  c  c | } 
  \hline
   
 \textbf{Benchmark} & Num. Inputs & $|$Training$|$ & Max. Training & $|$Testing$|$  & Max. Testing & Scale & Prediction & Error \% & Performance Gain\\ 
 \hline
 

Stateless  & \multirow{2}{8.0em}{12685.4 (+/- 703.9)} & \multirow{2}{8.0em}{3702.8 (+/- 1067.5)} & \multirow{2}{6.0em}{101 (+/- 0.0)} & \multirow{2}{8.0em}{8984.6 (+/- 1550.6)} & \multirow{2}{6.0em}{101 (+/- 0.0)} & \multirow{2}{6.0em}{5.9 (+/- 3.7)} & \multirow{2}{8.0em}{134.1 (+/- 18.4)} & \multirow{2}{7.0em}{32.8 (+/- 18.2)} & \multirow{2}{6.0em}{2,300,058} \\
Auth  &   &     &    &     &    &   &    &    & \\    
\hline
Jetty & \multirow{2}{8.0em}{13934.4 (+/- 185.0)} & \multirow{2}{6.0em}{532 (+/- 49.1)} & \multirow{2}{6.0em}{27.6 (+/- 2.2)} & \multirow{2}{8.0em}{13407.6 (+/- 229.8)} & \multirow{2}{6.0em}{54 (+/- 7.6)} & \multirow{2}{6.0em}{7.4 (+/- 1.8)} & \multirow{2}{6.0em}{75.1 (+/- 19.9)} & \multirow{2}{7.0em}{39.0 (+/- 43.7)} & \multirow{2}{6.0em}{1,241,133} \\ 
Safe &   &     &    &     &    &   &    &    & \\    
\hline

Jetty & \multirow{2}{8.0em}{12649.4 (+/- 620.6)} & \multirow{2}{8.0em}{6403.2 (+/- 470.83)} & \multirow{2}{6.0em}{51.4 (+/- 1.3)} & \multirow{2}{8.0em}{6250.2 (+/- 334.9)} & \multirow{2}{6.0em}{52 (+/- 0.0)} & \multirow{2}{6.0em}{10.8 (+/- 8.0)} & \multirow{2}{6.0em}{85.9 (+/- 61.4)} & \multirow{2}{7.0em}{65.2 (+/- 118.0)} & \multirow{2}{6.0em}{80,741} \\ 
 &   &     &    &     &    &   &    &    & \\    
\hline
Orientdb & \multirow{2}{8.0em}{13963 (+/- 556.6)} & \multirow{2}{8.0em}{5605.8 (+/- 1442.3)} & \multirow{2}{6.0em}{47 (+/- 0.0)} & \multirow{2}{8.0em}{8359.2 (+/- 1741.4)} & \multirow{2}{6.0em}{47 (+/- 0.0)} & \multirow{2}{8.0em}{13.63 (+/- 10.7)} & \multirow{2}{8.0em}{137.8 (+/- 73.5)} & \multirow{2}{7.0em}{193.2 (+/- 156.3)} & \multirow{2}{6.0em}{234,058} \\ 
 &   &     &    &     &    &   &    &    & \\    
\hline

Picketbox & \multirow{2}{8.0em}{13527.8 (+/- 771.0)} & \multirow{2}{8.0em}{7479 (+/- 1458.6)} & \multirow{2}{6.0em}{30.8 (+/- 0.45)} & \multirow{2}{8.0em}{6050.4 (+/- 994.6)} & \multirow{2}{6.0em}{31 (+/- 0.0)} & \multirow{2}{6.0em}{7.32 (+/- 2.4)} & \multirow{2}{6.0em}{72.1 (+/- 11.3)} & \multirow{2}{7.0em}{132.7 (+/- 36.4)} & \multirow{2}{6.0em}{193,613} \\ 
 &   &     &    &     &    &   &    &    & \\    
\hline
Spring Security & \multirow{2}{8.0em}{13807.6 (+/- 324.3)} & \multirow{2}{8.0em}{10419 (+/- 2855.2)} & \multirow{2}{6.0em}{149 (+/- 0.0)} & \multirow{2}{8.0em}{3328 (+/- 2472.5)} & \multirow{2}{6.0em}{149 (+/- 0.0)} & \multirow{2}{6.0em}{38.2 (+/- 18.6)} & \multirow{2}{8.0em}{333.7 (+/- 142.2)} & \multirow{2}{7.0em}{124.0 (+/- 95.4)} & \multirow{2}{6.0em}{479,232} \\ 
 &   &     &    &     &    &   &    &    & \\
\hline

Tourplanner & \multirow{2}{8.0em}{7465.2 (+/- 153.1)} & \multirow{2}{8.0em}{3205 (+/- 1805.1)} & \multirow{2}{8.0em}{341.2 (+/- 37.4)} & \multirow{2}{8.0em}{4290 (+/- 1830.0)} & \multirow{2}{8.0em}{380.8 (+/- 28.0)} & \multirow{2}{6.0em}{79.6 (+/- 83.7)} & \multirow{2}{8.0em}{784.7 (+/- 547.3)} & \multirow{2}{7.0em}{106.1 (+/- 160.8)} & \multirow{2}{6.0em}{1,434,977} \\ 
 &   &     &    &     &    &   &    &    & \\
\hline

Apache Ftpserver & \multirow{2}{8.0em}{13347.4 (+/- 178.7)} & \multirow{2}{8.0em}{5245 (+/- 4188.7)} & \multirow{2}{8.0em}{143.2 (+/- 33.5)} & \multirow{2}{8.0em}{8125.4 (+/- 4195.3)} & \multirow{2}{6.0em}{175 (+/- 18.6)} & \multirow{2}{6.0em}{25.2 (+/- 10.4)} & \multirow{2}{8.0em}{255.1 (+/- 78.4)} & \multirow{2}{7.0em}{45.793 (+/- 28.1)} & \multirow{2}{6.0em}{1,674,774,946} \\ 
Salted  &   &     &    &     &    &   &    &    & \\
\hline

Apache Ftpserver & \multirow{2}{10.0em}{10295.8 (+/- 1278.7)} & \multirow{2}{8.0em}{504.4 (+/- 103.7)} & \multirow{2}{6.0em}{53 (+/- 0.0)} & \multirow{2}{8.0em}{9793.4 (+/- 1264.5)} & \multirow{2}{6.0em}{53 (+/- 0.0)} & \multirow{2}{6.0em}{15.3 (+/- 2.0)} & \multirow{2}{8.0em}{162.6 (+/- 18.3)} & \multirow{2}{7.0em}{206.8 (+/- 34.5)} & \multirow{2}{6.0em}{571,164} \\ 
Stringutils   &   &     &    &     &    &   &    &    & \\
\hline

 \end{tabular}
 }
 \label{tab-RQ3}
\end{table*}

%% file: Sections/discussion.tex
\section{Discussion}
\label{sec:discussion}

The choice of extreme value theory over other potential approaches like machine learning-based extrapolation was deliberate. EVT provides a strong theoretical foundation specifically designed for modeling extreme events and tail behavior. Machine learning approaches, while flexible, typically focus on modeling the average case behavior rather than extremes, and may require significantly more training data to make reliable predictions about rare events. 
We focused our baseline comparisons on classical statistical methods (Markov, Chebyshev, Bayes factor) as they provide theoretical bounds with clear probabilistic interpretations. While other approaches like linear regression or more sophisticated time series models could be considered, they generally make stronger assumptions about the underlying distribution and may not be well-suited for modeling extreme events.

\replaced{\noindent \textbf{Limitation.} Our EVT model of differential fuzzing is limited to the specific program under test and the ongoing fuzzing campaign.
The model's purpose is not to predict a program's absolute worst-case difference, but to predict the worst-case difference likely to be discovered by the current fuzzing.}{Another important consideration is that our statistical guarantees apply to the fuzzing process itself, not directly to the programs being fuzzed. While we can predict with confidence the expected maximum cost differences that will be found through continued fuzzing, we cannot make absolute claims about the maximum possible cost differences that could exist in the program. Our predictions are bounded by what the fuzzing process is capable of discovering within its search space and mutation strategies.}

\added{\noindent \textit{Differential Metrics.}}
\added{Our framework supports differential metrics that are quantitative, ordinal, and have a sufficiently rich value space in its upper tail. This makes our approach broadly applicable to a range of common differential metrics beyond Java bytecode counts. For instance, it generalizes directly to performance testing (measuring execution time differences in microseconds) and resource consumption analysis (e.g., peak memory usage differences in bytes). 
In essence, any scenario where the ``worst-case" is characterized by the magnitude of a numeric difference falls within the ideal application scope of our method. 
}

\added{The efficacy of our approach degrades as the differential metric becomes more discrete. In the extreme case of a binary metric (e.g., passed vs. failed), the concept of an extreme magnitude collapses.  In this scenario, our EVT framework pivots from modeling how large the next difference will be to modeling the waiting time until the next (failed) event. This makes the problem conceptually similar to prior works that use Bernoulli or Poisson statistics to model bug discovery rates. However, a key distinction remains: while prior works bound the average discovery rate, our framework would model the tail distribution of the inter-arrival times between events.}

\replaced{\noindent \textit{Non-i.i.d. Setting.} Another core assumption of classical EVT is that samples are independent and identically distributed (i.i.d.). This assumption is violated in a graybox fuzzing setting. We mitigate this challenge in two ways. First, we hypothesize that after an initial warm-up phase, the fuzzing process enters a relatively stationary state, where the underlying distribution of observed differences becomes more stable. This addresses the ``identically distributed" aspect. Second, to handle the lack of independence, we employ a block bootstrapping technique based on recent work for dependent data~\cite{hrba2022bootstrapping}. This method resamples blocks of consecutive observations rather than individual points, preserving the local dependency structure within the samples. While this is a practical mitigation rather than a perfect theoretical fix, it is a standard approach~\cite{lahiri2013resampling,32a59ea1-164d-3469-8f79-e8540ef19dea,gilda2024tsbootstrapenhancingtimeseries} for applying statistical models to dependent data where true independence cannot be guaranteed.}{One key assumption in the statistics of extreme value is that the samples are independent of each other. However, graybox fuzzing may violate this assumption. To overcome this challenge, we used Bootstrapping~\cite{hrba2022bootstrapping} to eliminate the dependency between variables. We also use longer fuzzing durations to ensure that the process has become stationary to apply extreme value theory.}
\deleted{Our approach assumes that both training and testing data follow the same underlying distribution, which allows us to extrapolate from training samples to predict testing behavior. This assumption is reasonable since both phases use the same fuzzing process, just at different points in time.}

\added{\noindent \textit{Overheads.}}
\added{The overhead of our method is primarily related to bootstrapping (Algorithm 3), while the exponentiality test (Algorithm 2) is a lightweight statistical test. The bootstrapping step can take up to 6 minutes to complete per fuzzing campaign. However, this cost is easily justified. The expensive inference is a one-time cost that occurs only once per campaign, right before making a stopping decision. As our results show (Table IV), this one-time cost is far outweighed by the significant savings from early termination.}

\noindent \textbf{Threats to Validity.} 
To address \textit{internal validity} concerns and account for the stochastic nature of fuzzing, we adhered to best practices outlined in previous work \cite{Arcuri2014AssessFuzzing,Klees2018EvaluateFuzzing} for rigorous fuzzing evaluation.
In particular, we repeated the experiments $30$ times, showed the error margins, performed experiments with multiple seed inputs, and considered not only the final results, but also the temporal development. An area of weakness that requires further research is the relatively high variance in the outcomes of EVT. \added{We chose Exponential and Poisson Process as they are direct and principled applications of the Peaks-Over-Threshold methodology in EVT. The occurrences of events exceeding a high threshold can be modeled as a Poisson Process. Similarly, the size of these excesses (how much larger they are than the threshold) is theoretically modeled by the Exponential distribution.}

\input{Sections/Table_Ex1}

To address \textit{external validity} concerns, we evaluated our approach on a diverse set of benchmarks ranging from small micro-benchmarks to large real-world Java applications. Our subjects included programs from established benchmark suites like \blazer~\cite{antonopoulos2017decomposition}, \themis~\cite{DBLP:conf/ccs/ChenFD17}, and \diffuzz~\cite{8812124}, as well as widely-used libraries such as Spring Security and Apache FtpServer. This variety helps demonstrate that our EVT-based predictions generalize across different program sizes and complexity levels. 
\added{To show the applicability beyond \textsc{DifFuzz}, we explore the use of another differential fuzzing framework, called QFuzz~\cite{noller2021qfuzz}. Instead of byte-code cost differences, we leverage our EVT approach to predict the number of clusters inferred by QFuzz. We consider the leak set and password matching programs and run QFuzz on each benchmark for 30 minutes (repeated 5 times with different seeds). The results are presented in Table~\ref{table:EX1}. The average error in using the proposed EVT-based stopping criteria for QFuzz is below 1\%.
}

%% file: Sections/Table_Ex1.tex
\begin{table}
    \centering
    \caption{\added{EVT-based prediction of clustering in \textsc{QFuzz}~\cite{noller2021qfuzz}}}
    \resizebox{0.49\textwidth}{!}{%
        \begin{tabu}{|c c c c c c |}
        \hline
        Benchmark & $|$Training$|$ & $|$Testing$|$ & Ground Truth & Prediction & Error\%\\
        \hline
        
        Leaky (1) & 182 (+/-46.23) & 1618 (+/-46.23) & 13 (+/-0) & 12.96 (+/-0) & -0.27 (+/-0.02) \\
        \hline
        
        Leaky (2) & 893.25 (+/-169.56) & 906.75 (+/-169.56) & 16.99 (+/-0.01) & 17.1 (+/-0.27) & 0.65 (+/-1.54) \\
        \hline
        
        Leaky (3) & 1701 (+/-113.2) & 99 (+/-113.2) & 20.89 (+/-0.03) & 20.9 (+/-0.12) & 0.07 (+/-0.52) \\
        \hline
        
        Leaky (4) & 1756.4 (+/-36.69) & 43.6 (+/-36.69) & 24.5 (+/-0.1) & 24.44 (+/-0.11) & -0.25 (+/-0.06)\\
        \hline
        
        Leaky (5) & 1749.6 (+/-29.68) & 50.4 (+/-29.68) & 27.66 (+/-0.38) & 27.6 (+/-0.38) & -0.22 (+/-0.02) \\
        \hline

        Password Matching & 1667 (+/-146.14) & 133 (+/-146.14) & 16.81 (+/-0.22) & 16.7 (+/-0.27) & -0.65 (+/-0.32) \\
        \hline
        
        \end{tabu}
    }
    \label{table:EX1}
    \vspace{-1.0 em}
\end{table}

        
        
        
        
        
        

%% file: Sections/conclusion.tex
\section{Conclusion}
\label{sec:conclusion}
In this paper, we explored the application of Extreme Value Theory (EVT) to provide statistical guarantees on the worst-case divergence in differential fuzzing. We adapted EVT to model the maximum cost differences of differential fuzzing and analyzed the tail distribution of these differences. Through extensive experiments on real-world Java libraries and web servers, we demonstrate that EVT can effectively predict the maximum cost differences in side-channel analysis. It also outperformed the baseline statistical methods. Finally, we showed EVT-enabled differential fuzzing can provide significant performance gains through early termination.

There are multiple interesting directions for future work. First, we plan to extend our approach to other differential testing domains, such as machine learning models and libraries. Second, we plan to explore the optimal hyperparameters for EVT distributions and the optimal termination condition for differential fuzzing campaigns.


\noindent \textbf{Acknowledgment.} The authors thank the anonymous ASE reviewers for their time and invaluable feedback to improve this work. 
This project has been supported by NSF under Grant No. CNS-2230060 and CNS-2527657.